\documentclass{aa}  
\PassOptionsToPackage{disable}{showkeys}
\usepackage{graphicx}
\usepackage{txfonts}
\usepackage{longtable}
\usepackage{hyperref}

\begin{document}

   \title{Broadband Polarized Radio Emission Detected from Starlink Satellites Below 100 MHz with NenuFAR}

   \author{X. Zhang\inst{1}
            \and
            P. Zarka\inst{1,2}
            \and 
            C. Viou\inst{2}
            \and
            A. Loh\inst{1}
            \and
            C. G. Bassa\inst{3}
            \and
            Q. Duchene\inst{1}
            \and
            C. Tasse\inst{4}
            \and
            J-M. Grießmeier\inst{5,2}
            \and
            J. D. Turner\inst{6}
            \and
            O. Ulyanov\inst{7}
            \and
            L. V. E. Koopmans\inst{8}
            \and
            F. Mertens\inst{4,8}
            \and
            V. Zakharenko\inst{7}
            \and
            C. Briand\inst{1,2}
            \and 
            B. Cecconi\inst{1,2}
            \and
            R. Vermeulen\inst{9}
            \and
            O. Konovalenko\inst{7}
            \and
            J. Girard\inst{1}
            \and
            S. Corbel\inst{10}
          }

   \institute{LIRA, Observatoire de Paris, Université PSL, Sorbonne Université, Université Paris Cité, CY Cergy Paris Université, CNRS, 92190 Meudon, France\\
        \email{Xiang.Zhang@obspm.fr}
        \and
        Observatoire Radioastronomique de Nançay (ORN), Observatoire de Paris, Université PSL, Univ Orléans, CNRS, 18330 Nançay, France
        \and
        ASTRON, Netherlands Institute for Radio Astronomy, Oude Hoogeveensedijk 4, 7991 PD Dwingeloo, The Netherlands
        \and
        LUX, Observatoire de Paris, Université PSL, Sorbonne Université, CNRS, 75014 Paris, France
        \and
        LPC2E, OSUC, Univ Orleans, CNRS, CNES, Observatoire de Paris, F-45071 Orleans, France
        \and
        Department of Astronomy and Carl Sagan Institute, Cornell University, Ithaca, NY, USA
        \and
        IRA NASU, Institute of radio astronomy of NAS of Ukraine, Kharkiv, Ukraine
        \and
        Kapteyn Astronomical Institute, University of Groningen, PO Box 800, 9700 AV Groningen, The Netherlands
        \and
        LOFAR ERIC, Oude Hoogeveensedijk 4, Dwingeloo, The Netherlands
        \and
        Université Paris Cité and Université Paris Saclay, CEA, CNRS, AIM, 91190 Gif-sur-Yvette, France
        }

   \date{Received September 15, 1996; accepted March 16, 1997}

 
  \abstract
   {}
   {This study evaluates the impact of Starlink satellites on low-frequency radio astronomy below 100 MHz, focusing on challenges on data processing and scientific goals.}
   {We conducted 40 hours of imaging observations using NenuFAR, in the 30.8–78.3 MHz range. Observations included both targeted tracking of specific satellites based on orbital predictions and untargeted searches focused on high-elevation regions of the sky. Images in total intensity and polarimetry were obtained, and full Stokes dynamic spectra were generated for several hundred directions within the Field of View. Detected signals were cross-matched with satellite orbital data to confirm satellite associations. Detailed analyses of the observed spectra, polarization, and temporal characteristics were performed to investigate the origin and properties of the detected emissions.}
   {We detected broadband emissions from Starlink satellites, predominantly between 54–66 MHz, with flux densities exceeding 500 Jy. These signals are highly polarized and unlikely to originate from ground-based RFI or reflected astronomical sources. Instead, they are likely intrinsic to the satellites, with distinct differences in emission properties observed between satellite generations. These findings highlight significant challenges to data processing and scientific discoveries at these low frequencies, emphasizing the need for effective mitigation strategies, particularly through collaboration between astronomers and satellite operators.}
   {}


    \titlerunning{Starlink with NenuFAR}
    \authorrunning{X. Zhang et al}

   \maketitle
%

\section{Introduction}

Satellite contamination has long been a significant concern for astronomical observations across various wavelengths. In optical astronomy, satellites orbiting Earth reflect sunlight, creating bright streaks or flares in telescope images. A notable example is the "Iridium flare," which can be as bright as magnitude $-8$ \citep{2020A&A...636A.121H}. In radio astronomy, Global Navigation Satellite System (GNSS) satellites are known to contribute to radio frequency interference (RFI), probably through out-of-band emissions, harmonics, or intermodulation products \citep{wildemeersch2010radio}. Additionally, satellites residing in low Earth orbit (LEO) can reflect ground-based RFI back to Earth \citep{2020PASA...37...52P}. This widespread contamination affects multiple areas of astronomical research \citep{2020BAAS...52.0206W}, such as the study of transient phenomena, deep imaging surveys, spectroscopic studies, and the search for Near-Earth Objects (NEOs), thereby hindering our ability to observe and understand the universe accurately.

Over the past decade, the advent of satellite mega-constellations, especially SpaceX's Starlink, OneWeb, and Amazon's Project Kuiper, has significantly intensified concerns about satellite contamination in optical astronomy \citep{2020BAAS...52.0206W}. These large fleets, comprising thousands of satellites in LEO, are particularly problematic due to their sheer numbers, low altitudes, and potential high reflectivity \citep{2020arXiv200110952G}. Estimates indicate that at low elevations near twilight at intermediate latitudes, hundreds of satellites may be visible at once to naked-eye observers \citep{2020ApJ...892L..36M}. A study using ESO telescopes shows that this contamination is especially critical for wide-field surveys and long-exposure observations, where up to 3\% of the first and last hours of their nighttime observations are affected \citep{2020A&A...636A.121H}. Besides ground-based telescopes, space-based telescopes are affected as well. As of 2021, approximately 2.7\% of the typical individual exposure images taken with the Hubble Space Telescope are crossed by satellites \citep{2023NatAs...7..262K}.

In addition to their impact on optical astronomy, Starlink satellites have raised significant concerns within the radio astronomy community. These concerns are particularly acute for low-frequency radio astronomers due to the satellites’ potential to generate radio frequency interference (RFI). Instruments such as the Low-Frequency Array (LOFAR) \citep{2013A&A...556A...2V} have detected unintentional radio emissions from these satellites in total intensity and beamformed mode, including broadband emissions between 110 and 188MHz and narrowband emissions at 125, 135, 143.05, 150, and 175MHz \citep{2023A&A...676A..75D}, as well as emissions in the 40–70~MHz range \citep{2024A&A...689L..10B}. Similarly, observations with the Engineering Development Array version 2 (EDA2) \citep{2022JATIS...8a1010W}, a SKA-Low prototype station located at the site of the future SKA-Low facility, have detected Starlink satellites at 137.5~MHz and 159.4~MHz, with intensities reaching up to $10^{6}$~Jy\,beam$^{-1}$ \citep{2023A&A...678L...6G}. A recent study using the prototype SKA-Low stations detected various satellite signals between 137 and 400 MHz, but did not find emissions in radio astronomy-protected frequency bands \citep{grigg2025enhanced}. Additionally, transient RFI from non-ground-based sources, including aircraft and satellites, has been identified as a challenge for 21-cm cosmology experiments with LOFAR-AARTFAAC \citep{2024A&A...681A..71G}. These emissions contaminate the radio spectrum used by astronomers to observe faint cosmic signals, posing a significant challenge for radio telescopes that require radio-quiet environments.

Given the above, we propose to further investigate the potential contamination from Starlink satellites at frequencies below 100 MHz. Observations at these low frequencies are essential for advancing our knowledge in multiple scientific fields, notably stellar and planetary radio emissions \citep{2004ApJ...612..511L, 2015aska.confE.120Z, zarka2024star}, Cosmic Dawn and the Epoch of Reionization (EoR) \citep{2012RPPh...75h6901P, 2013ExA....36..235M, 2015aska.confE...1K, munshi2024first}, and pulsar science \citep{2011A&A...530A..80S, 2016A&A...585A.128K, bondonneau2021pulsars}. Consequently, estimating the level of contamination caused by Starlink at these frequencies is crucial for understanding their impact on low-frequency radio astronomy, particularly for SKA-Low \citep{labate2022highlights}, and for developing effective mitigation strategies.

In this paper, we present a search for radio emissions below 100\,MHz using imaging data from the New Extension in Nançay Upgrading LOFAR (\textit{NenuFAR}; see \citealt{2012sf2a.conf..687Z, zarka2020low}), aiming to assess the contamination impact in both total intensity and polarization. Section~\ref{sec:obs} outlines the details of our observational campaign, while Section~\ref{sec:process} describes the data processing strategy implemented to extract the signals. Detection results, including imaging, spectral and polarimetric properties, are presented in Section~\ref{sec:results}. In Section~\ref{sec:discuss}, we explore the origins of the detected emissions and their broader implications for low-frequency radio astronomy. Finally, Section~\ref{sec:conclusion} summarizes our findings and discusses possible mitigation strategies to address the challenges posed by satellite mega-constellations.

\section{Observations}
\label{sec:obs}

Our observations were carried out with NenuFAR, a low-frequency phased array optimized for the 10--85\,MHz frequency range. The array consists of clusters of dipole antennas, each cluster referred to as a Mini-Array (MA), with each MA containing 19 dipole antennas. By mid-2024, NenuFAR comprises 80 MAs distributed within a core area of 400 meters, along with an additional 4 remote MAs located up to 3\,km from the core.

The observations were conducted between June and July 2024, adding up to 40 hours. Both imaging and beamformed modes were employed simultaneously; however, this paper focuses solely on the imaging observations. Due to correlator limitations, imaging observations were conducted using a subset of NenuFAR's full bandwidth, covering 30.8 - 78.3 MHz. This frequency range was chosen for its superior data quality, as lower frequencies are persistently affected by radio frequency interference (RFI) from ground-based transmitters, which is consistently observed in system monitoring. Additionally, ionospheric turbulence further degrades data quality at these frequencies, making them less suitable for our study. Higher frequencies, on the other hand, suffer from reduced antenna response. Imaging data were recorded with a time resolution of 1 second to accommodate the rapid motion of satellites, which traverse multiple PSF widths per second (e.g., >1°/s compared to a $\sim$0.5° PSF at 40 MHz). A frequency resolution of 97.5\,kHz was selected to balance data rate constraints and support the detection of potential narrow-band emissions. All core and remote MAs were used during these observations, with analog phasing of the MAs providing a limited instantaneous field of view (FoV). Table~\ref{tab:obs_para} summarizes the observational parameters.

Our campaign employed two approaches: targeted observations and untargeted searches. For targeted observations, we selected satellites based on the brightest detections made by LOFAR \citep{2023A&A...676A..75D} or early detections from our NenuFAR campaign. Using Two-Line-Element (TLE) data\footnote{\url{https://www.space-track.org/}} \citep{vallado2012two}, we calculated the predicted orbital paths of specific satellites as they passed over NenuFAR. As NenuFAR cannot track satellites directly, observations were scheduled to follow astronomical sources at the celestial coordinates (Right Ascension and Declination) of each satellite’s maximum elevation point, ensuring the satellite passed through the FoV. Each targeted observation was conducted in tracking mode and lasted several tens of minutes, centered on the satellite's passage time.

For untargeted searches, we tracked astronomical sources passing by the zenith to optimize NenuFAR's sensitivity. We did not observe the zenith in transit mode, as imaging observations in transit mode were not supported by NenuFAR at the time of our campaign. Each untargeted search typically spanned several hours.

Table~\ref{table:obs_dates} lists the dates and times of our observations, along with notes specifying the observation approach used and, for targeted observations, the specific satellite being tracked. To ensure accurate calibration, a 10-minute observation of Cygnus A was performed before or after each observation session. 

\begin{table}
\caption{Observational Parameters for the NenuFAR Starlink Campaign.}
\label{tab:obs_para}
\centering
\begin{tabular}{c c c}
\hline\hline
Parameter & Value & Units \\
\hline
Array Configuration & Core + Remote MAs & \\
Frequency Range & 30.8 to 78.3 & MHz \\
Frequency Resolution & 97.5 & kHz \\
Temporal Resolution & 1 & Second \\
Field of View (FoV) \tablefootmark{a} & 24 to 10 & Degrees \\
\hline
\end{tabular}
\tablefoot{
    \tablefoottext{a}{The FoV is calculated as the full width at half maximum (FWHM) of NenuFAR at the minimum (30.8 MHz) and maximum (78.3 MHz) observed frequencies.}
    }
\end{table}

\begin{table*}
\caption{Observation Dates and Times for Starlink Satellites. The table lists the dates, times, and observation notes of our campaign. Observations labeled “untargeted search around zenith” were conducted by tracking an astronomical source passing overhead. Observations labeled with specific targets (e.g., “Observing STARLINK-3647 and STARLINK-3657”) involved tracking an astronomical radio source located at the point of highest elevation along the satellites’ paths across the sky, as illustrated in Figure \ref{Fig:path}. }             
\label{table:obs_dates}      
\centering                          
\begin{tabular}{l l l l l}        
\hline\hline                 
Start Date & Start Time (UT) & End Date & End Time (UT) & Observation notes \\    
\hline                        
2023-04-18 & 08:00 & 2023-04-18 & 10:00 & Untargeted search around zenith \\
2023-07-17 & 10:00 & 2023-07-17 & 11:00 & Observing STARLINK-3647 and STARLINK-3657 \\
2024-06-06 & 16:00 & 2024-06-06 & 17:00 & Observing STARLINK-3647 and STARLINK-3657 \\
2024-06-20 & 03:00 & 2024-06-20 & 05:00 & Untargeted search around zenith \\
2024-06-20 & 13:00 & 2024-06-20 & 21:00 & Untargeted search around zenith \\
2024-06-21 & 03:00 & 2024-06-21 & 06:00 & Untargeted search around zenith \\
2024-06-21 & 17:00 & 2024-06-21 & 19:00 & Untargeted search around zenith \\
2024-06-22 & 01:00 & 2024-06-22 & 04:00 & Untargeted search around zenith \\
2024-06-22 & 17:00 & 2024-06-22 & 19:00 & Untargeted search around zenith \\
2024-06-23 & 17:00 & 2024-06-23 & 19:00 & Untargeted search around zenith \\
2024-06-24 & 03:00 & 2024-06-24 & 06:00 & Observing STARLINK-31034 \\
2024-06-25 & 03:00 & 2024-06-25 & 08:00 & Untargeted search around zenith \\
2024-06-30 & 23:00 & 2024-07-01 & 01:00 & Nighttime untargeted search around zenith \\
2024-07-03 & 00:00 & 2024-07-03 & 01:00 & Nighttime untargeted search around zenith \\
2024-07-03 & 04:50 & 2024-07-03 & 05:20 & Observing the International Space Station \\
2024-07-04 & 00:00 & 2024-07-04 & 02:00 & Nighttime untargeted search around zenith \\
2024-07-04 & 04:00 & 2024-07-04 & 04:30 & Observing the International Space Station \\
\hline                                   
\end{tabular}
\end{table*}

\section{Data Processing}
\label{sec:process}

Our data processing method follows the standard NenuFAR exoplanet imaging pipeline (Zhang et al., in prep). A combination of NenuFAR and LOFAR tools were used throughout the data processing, including Nenupy \citep{alan_loh_2020_4279405}, Nenucal\footnote{https://gitlab.com/flomertens/nenucal-cd} \citep{2024A&A...681A..62M}, DP3 \citep{2018ascl.soft04003V}, AOFlagger \citep{2012A&A...539A..95O}, DDFacet \citep{2018A&A...611A..87T}, KillMS \citep{2023ascl.soft05005T}, DynspecMS 
\citep{Tasse2025}, 
and WSClean \citep{offringa-wsclean-2014}.

The data processing began with flagging low-quality visibilities to remove contamination from instrumental issues and RFI. MAs affected by temporary instrumental failures were identified through a test-run calibration using Cygnus A, in which the calibration solutions for each MA were examined. Visibilities contaminated by RFI were flagged using a predefined RFI channel list and an automatic detection algorithm provided by AOFlagger.

After flagging, calibration was performed in two stages. The first stage involved an initial full-Stokes calibration using Cygnus A as a reference source to establish the absolute flux scale. The derived amplitude and phase calibration solutions were then applied to the observation field. In the second stage, a direction-dependent calibration was conducted to mitigate contamination from bright A-team sources and from the Sun.

\begin{figure}
\centering
\includegraphics[width=\columnwidth]{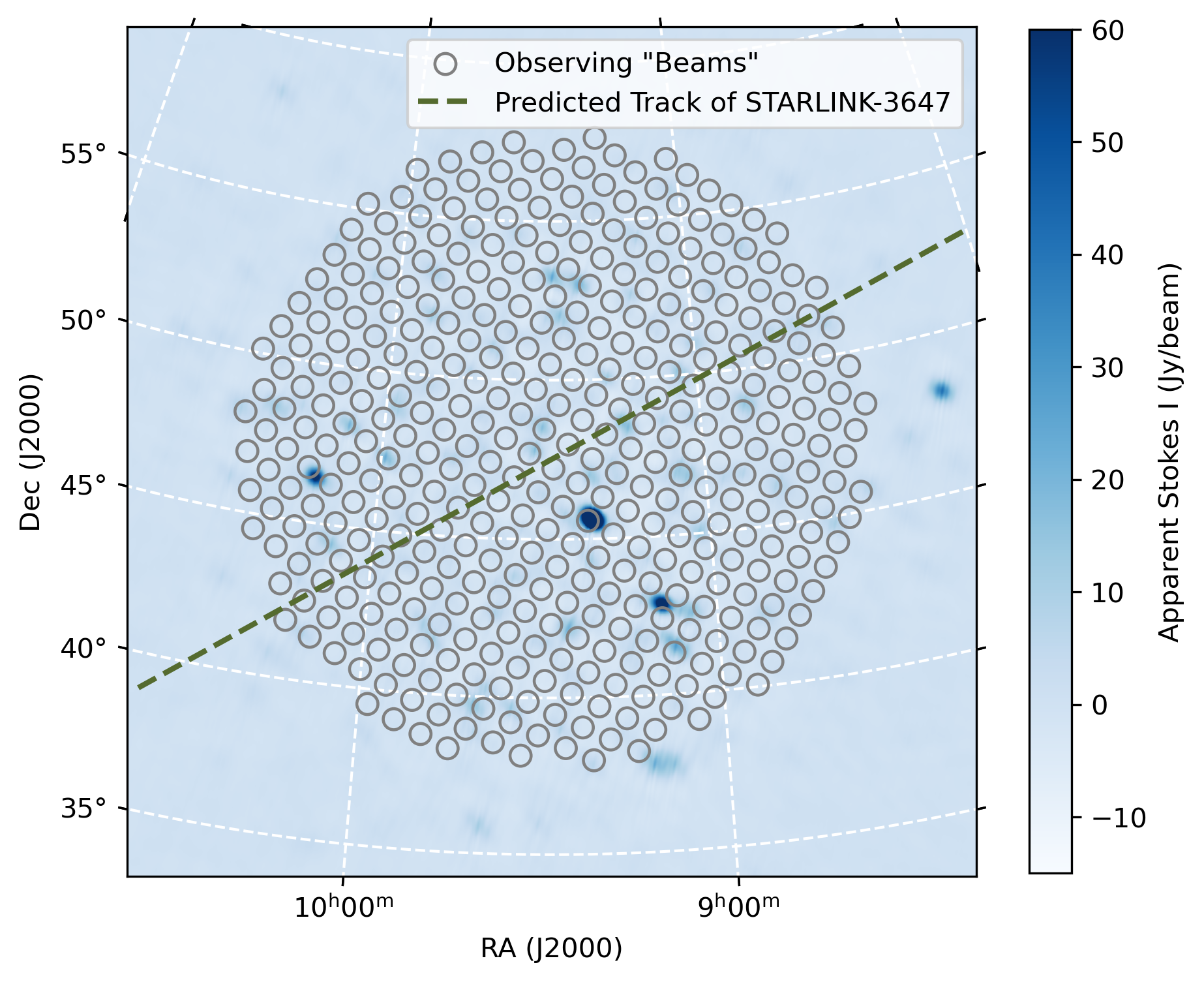}
\caption{Distribution of the dynamic spectra directions ("beams") within our observing FoV, overlaid on a NenuFAR broadband (30-78 MHz) Stokes I image of background astronomical radio sources. Each grey circle corresponds to a direction for the synthesized IQUV dynamic spectra. The dashed line represents the predicted path of STARLINK-3647 across the FoV from 2024-06-06T16:19:57 to 2024-06-06T16:20:31, calculated using TLE data.}
\label{Fig:path}
\end{figure}

After subtracting bright contaminating sources, IQUV dynamic spectra were generated for several hundred directions within the FoV. These spectra were derived from residual visibilities after constructing a sky model through imaging and removing radio continuum sources. The residual visibilities were then phased to a few hundred directions and summed to produce the final dynamic spectra (Tasse et al. submitted). Although the observations were conducted in imaging mode, these several hundred directions can effectively be treated as individual "beams". These pencil-beams were evenly spaced one degree apart, matching NenuFAR's resolution at 40\,MHz, and covered a 20-degree field across the sky. Figure~\ref{Fig:path} illustrates an example of the beams within the FoV, showing the predicted path of STARLINK-3647 passing through multiple beams during a targeted observation.

We searched for bursts in Stokes I and V dynamic spectra to investigate potential emissions from Starlink satellites. At low frequencies, linear polarization is highly susceptible to Faraday rotation \citep{2005A&A...441.1217B}, making it less reliable for burst detection. In contrast, circular polarization (Stokes V) is widely used in satellite communication systems \citep{toh2003understanding}, providing a strong motivation to include it in our analysis. To enhance detection sensitivity, we convolved the dynamic spectra across multiple time and frequency resolutions. Given the rapid angular velocities of Starlink satellites, which can reach several degrees per second near the zenith, we applied time windows of 1, 2, and 4 seconds. The convolved frequency resolutions ranged from 240\,kHz to 15.36\,MHz. A source-finding threshold of 6\,$\sigma$ was used to identify significant bursts at each time-frequency scale.

To verify the association between the detected bursts and Starlink satellites, we used satellite positions predicted from TLE data\footnote{The TLE data were retrieved from \url{https://www.space-track.org/} using a database query, with constraints on the epoch range to match the time span of our observations.}. TLE data provides the orbital elements required to calculate the right ascension (RA) and declination (Dec) of a satellite at a specific epoch. For each night’s observations, we calculated the predicted times and positions of satellite passes across the FoV, identifying the beams and epochs where satellites would be closest to the telescope’s pointing direction. These predictions were then crossmatched with the bursts detected in the dynamic spectra. To account for potential inaccuracies in the TLE data, we allowed a time mismatch of up to 3 seconds between the predicted and observed detections, which corresponds to a spatial uncertainty of a few degrees for Starlink satellites. A satellite was classified as “detected” if a burst was found within the corresponding predicted time window and beam.

\section{Results}
\label{sec:results}

Over the course of 40 hours of observations, we detected radio signals from 134 Starlink satellite passes. These detections were largely from satellites that were not our primary targets. Among the targeted observations, STARLINK-31034 and the ISS were detected, while STARLINK-3647 and STARLINK-3657 were not. Note that 5 hours of data were excluded from the analysis due to adverse weather conditions and temporary instrumental issues, resulting in an effective detection rate of approximately one satellite pass every 15 minutes for a field of 20 degrees size. A comprehensive summary of these detections is provided in Table \ref{tab:passes}, including details such as the passing time, flux density, and peak frequency of the emissions.

\begin{figure*}
\centering
\includegraphics[width=\textwidth]{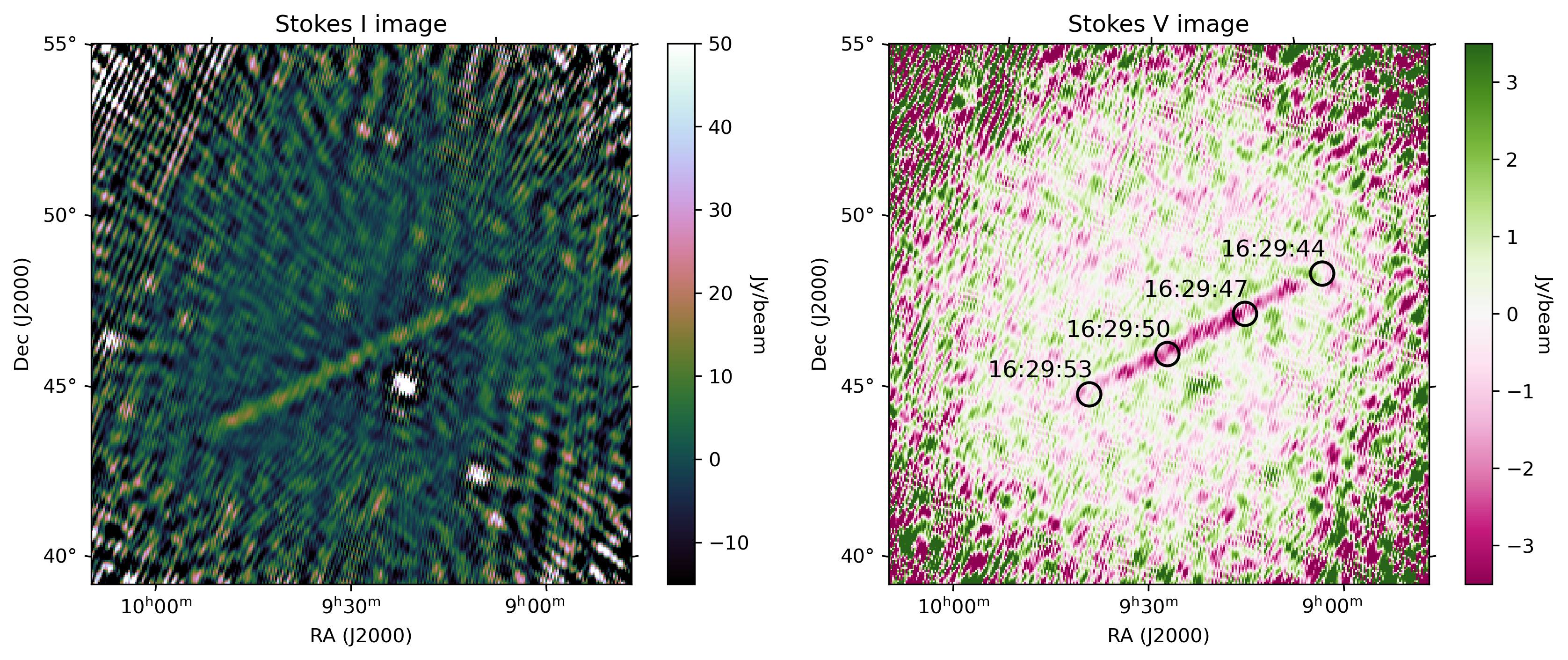}
\caption{Stokes I and Stokes V images produced with NenuFAR data, illustrating STARLINK-31034 passing through the FoV. The images were created using data collected between 54 and 66 MHz during the time period from 2024-06-06T16:29:46 to 2024-06-06T16:29:57. Black circles overlaid on the Stokes V image represent the predicted positions of the Starlink satellite, calculated using TLE data. The corresponding times for each position are annotated next to the circles, illustrating the satellite’s motion.}
\label{Fig:trail}
\end{figure*}

An example of a detected satellite pass is shown in Figure \ref{Fig:trail}, which highlights the trail left by STARLINK-31034 as it crossed NenuFAR’s FoV. Clear trails are visible in both Stokes I and V images. The predicted positions of the satellite, calculated using TLE data, are overlaid on the Stokes V image as black circles, and the satellite’s observed trajectory closely aligns with these predictions. Both images were generated using visibilities within the 54–66 MHz frequency range, corresponding to the peak of the broadband spectra of the satellites (see Section \ref{sec:continuum}). The data were integrated over a 15-second window around the time of detection, matching the satellite’s estimated passage duration. 

Within the two image panels, the signal-to-noise ratio (SNR) is significantly higher in the Stokes V image, indicating that the satellite emission is strongly polarized (see Section \ref{sec:polarized}). The high SNR is also caused by the lower noise levels in Stokes V (0.65 Jy/beam), as the sky background is predominantly unpolarized at low frequencies. In contrast, the Stokes I image displays a higher overall noise level (3.2 Jy/beam) due to confusion noise and PSF side-lobe, together with bright points corresponding to background astronomical radio sources.

\subsection{Continuum spectra}
\label{sec:continuum}
Visual inspection of the dynamic spectra revealed that Starlink satellite emissions are primarily concentrated between 54 and 66 MHz. Consequently, the flux density of the detected emissions was calculated by integrating over this frequency range (see Table \ref{tab:passes}).

\begin{figure*}
\centering
\includegraphics[width=\textwidth]{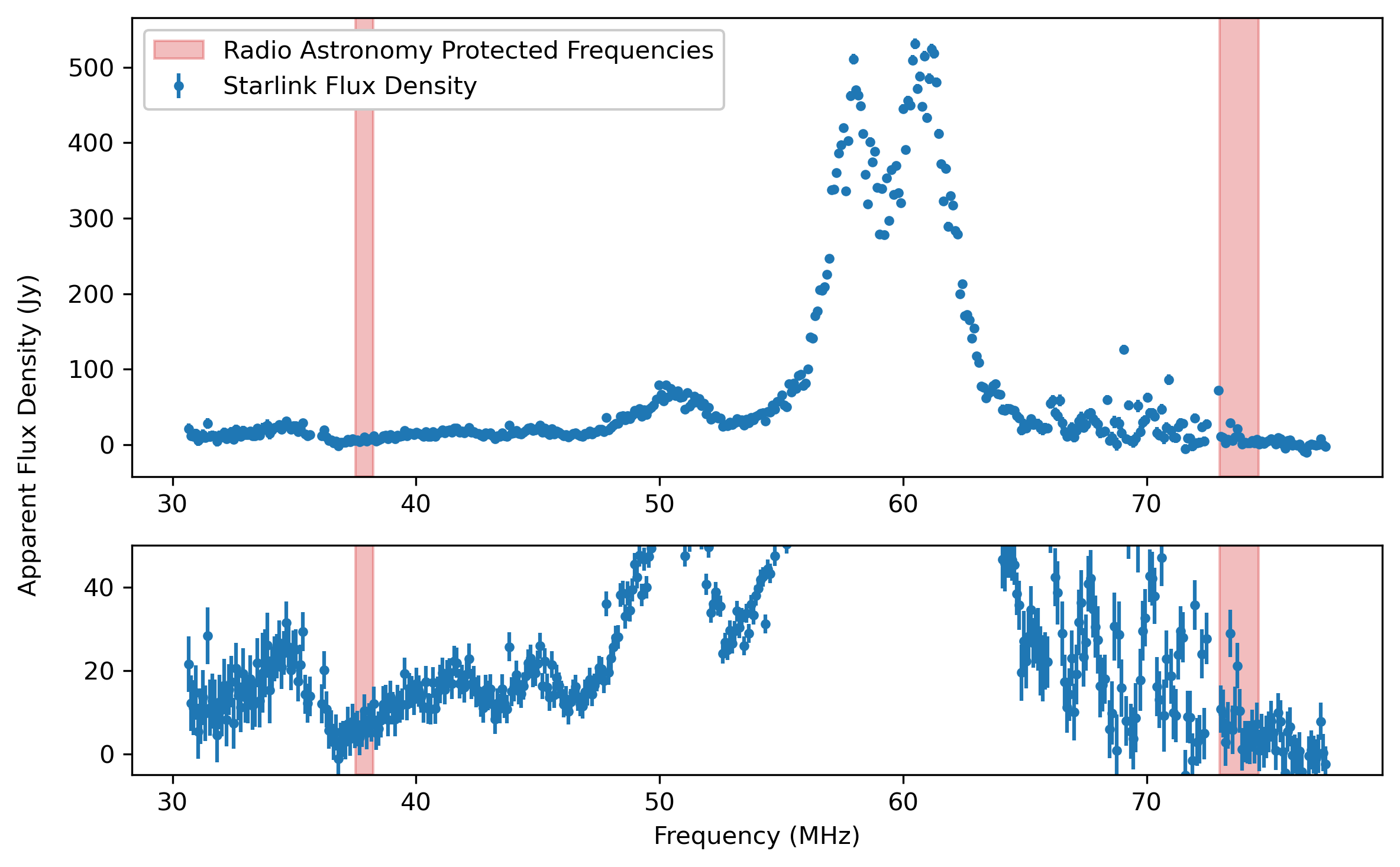}
\caption{Broadband Stokes I spectrum for STARLINK-31034, stacked along its trajectory across the NenuFAR FoV. The top panel shows the apparent flux density across the full NenuFAR observing frequency range, while the bottom panel provides a zoomed-in view with apparent flux density ranging from 0 to 50 Jy to highlight details of the spectrum outside the peak frequencies. In both panels, the radio astronomy-protected frequency bands (37.5–38.25 MHz and 73–74.6 MHz) are shaded in red.}
\label{Fig:spec}
\end{figure*}

Using this definition, the flux density of the detected satellite passes ranges from a few Jy to several hundred Jy. Notably, the brighter passes (> 100 Jy) exhibit a consistent pattern in their broadband Stokes I spectra: the satellite emissions are detectable within the broader range of 33–73 MHz, with a distinctive “double-peak” structure clearly visible between 54 and 66 MHz, as shown in Figure \ref{Fig:spec}. In this figure, each channel represents a 97.5 kHz frequency bin, and both the signal and noise were calculated by stacking multiple dynamic spectra along the satellite’s trajectory to enhance the SNR.

The actual flux density \footnote{In radio interferometry, the \textit{apparent flux density} refers to the measured signal before applying corrections for instrumental and observational effects, whereas the \textit{actual flux density} represents the intrinsic brightness of the source after applying necessary corrections. In our case, the actual flux density were obtained after applying the primary beam correction, which accounts for the telescope’s varying sensitivity across the FoV.} is expected to be higher than the apparent flux density shown in Figure \ref{Fig:spec} due to two key factors: (1) the apparent flux density was used to ensure a consistent noise level across the satellite’s path for stacking purposes, while the actual flux density would be higher after applying primary beam correction; and (2) the flux density estimation was based on dynamic spectra generated for fixed pointings, while the satellite moved through the beams, resulting in an averaged measurement.

A particularly concerning observation is the detection of satellite emissions within frequency bands specifically protected for radio astronomy, as defined in the Radio Regulations\footnote{\url{https://www.itu.int/pub/R-REG-RR}, accessed January 2025.} published by the Radiocommunication Sector of the International Telecommunication Union (ITU-R). Two such protected bands fall within our observational range: 37.5–38.25 MHz (continuum observations) and 73–74.6 MHz (solar wind observations and continuum observations). For the STARLINK-31034 pass, no significant emission was detected in the first protected band; however, the flux density exceeded 5 $\sigma$ between 73.29 and 73.39 MHz within the second band, with an additional tentative detection of 3.9 $\sigma$ between 73.39 and 73.48 MHz. This confirms that Starlink satellites are capable of producing detectable emissions within frequency ranges explicitly allocated to radio astronomy. However, these emissions are weak and only affect a limited region of the sky for a brief duration during each satellite pass.

\subsection{Polarized spectra}
\label{sec:polarized}
Unlike the consistency observed in total intensity, the polarization properties of the detected emissions vary significantly across satellite passes. For each pass, the polarization spectrum exhibits fluctuations as the satellite moves through the FoV. This variability is likely influenced by the satellite-telescope geometry: polarization is inherently direction-dependent, and the relative positioning of the satellite and telescope may affect the observed polarization characteristics.

\begin{figure*}
\centering
\includegraphics[width=\textwidth]{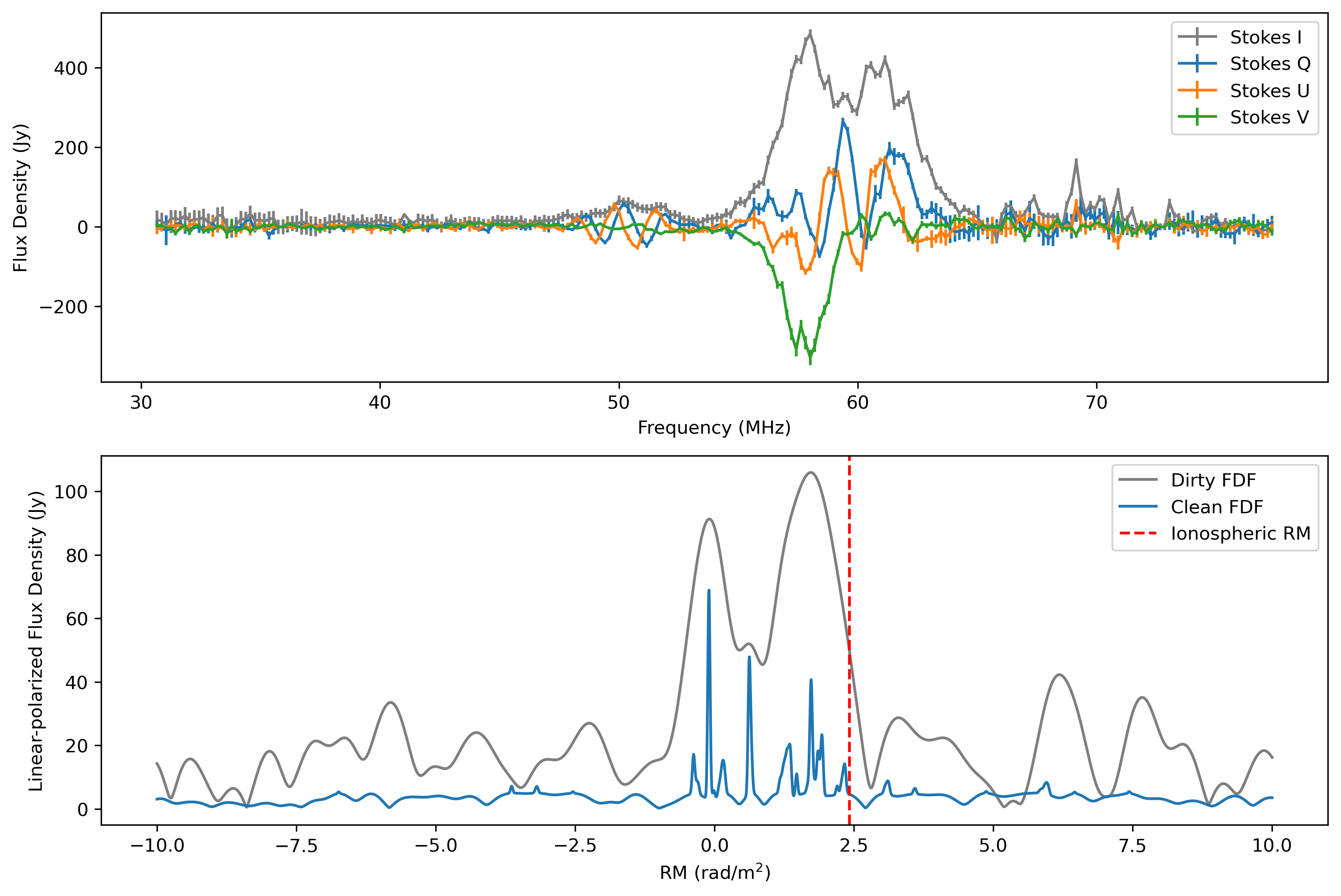}
\caption{Polarized spectra for STARLINK-31034, obtained from a single beam near the center of the FoV. The top panel displays the Stokes IQUV spectra of the satellite, while the bottom panel shows the result of RM synthesis on the linear polarization. The red vertical line in the bottom panel indicates the ionospheric RM at the time the IQUV spectra were collected, as estimated using GPS data.}
\label{Fig:IQUV}
\end{figure*}

Despite this variability, a notable feature common to all detected satellite passes is the high degree of fractional polarization, as illustrated in Table \ref{tab:passes}. Figure \ref{Fig:IQUV} shows full-Stokes spectra from a satellite path near the zenith. For this analysis, we used the IQUV dynamic spectra collected when the satellite was closest to the phase center. Analysis of these spectra indicates that the broadband emission was approximately 90\% polarized, comprising 68\% circular polarization and 22\% linear polarization\footnote{Although full-Stokes calibration was performed, residual instrumental polarization (i.e., leakage between Stokes parameters) is still present in NenuFAR images. However, these leakage effects are typically at the level of a few percent, thus significantly lower than the $\sim$90\% polarization observed from Starlink satellites.}.

In the linear polarization spectra, evidence of the Faraday effect was observed. Stokes Q and U spectra displayed sinusoidal patterns consistent with Faraday rotation. To quantify this effect, we performed Rotation Measure (RM) synthesis \citep{2005A&A...441.1217B} on the Stokes Q and U spectra and compared the resulting Faraday Dispersion Function (FDF) with ionospheric RM measurements derived from GPS data \citep{mevius2018rmextract}. The FDF revealed multiple peaks, including one near RM = 0 and others at nonzero RM values. The peak near zero RM is likely due to residual instrumental polarization (leakage), a known issue in low-frequency polarimetric imaging. Meanwhile, the presence of additional RM components suggests that the emission mechanism may not be stable, potentially indicating malfunctions or unintended behaviors within the satellite’s onboard systems. 

\section{Discussion}
\label{sec:discuss}

The detection of radio emissions from Starlink satellites raises several questions about the nature of these signals and their implications for low-frequency radio astronomy. Are these emissions the result of direct satellite transmissions, reflections of astronomical or terrestrial signals, or unintended intrinsic emissions from the satellites themselves? What physical mechanisms could underlie their observed frequency, polarization, and variability?

Beyond their origin, these signals prompt concerns about their impact on various areas of radio astronomy, including transient searches and non-transient studies such as spectroscopic and continuum surveys and cosmic magnetism. With the rapid expansion of satellite mega-constellations, what strategies can be developed to mitigate their influence and ensure the quality of astronomical data?

In this section, we delve into these questions, aiming to provide some groundwork for future investigations and foster collaboration between the astronomical community and satellite operators.

\subsection{Origin of Detected Emission from Starlink}

Given the constraints of international frequency allocations, it is improbable that the detected signals represent direct, intentional emissions from Starlink satellites. Starlink satellites primarily operate in higher frequency bands allocated for satellite communication, including the Ku-band (10.7--12.7\,GHz and 14.0--14.5\,GHz), Ka-band (17.7--20.2\,GHz and 27.5--30.0\,GHz), V-band (37.5--51.4\,GHz), and E-band (71.0--76.0\,GHz and 81.0--86.0\,GHz), which are used for data transmission and reception with ground stations and users. These operational frequencies, authorized by the Federal Communications Commission (FCC), are significantly higher than the 60\,MHz range in which our detections were made.\footnote{Details of the frequency allocations for Starlink satellites can be found in the FCC document DA-24-1193A1, available at https://docs.fcc.gov/public/attachments/DA-24-1193A1.pdf (accessed November 2024).}

This observation narrows the potential origins of the detected emissions to two main hypotheses: either the signals are reflections of radio signals by the satellites, or they are unintended emissions generated by the satellites themselves. 

\subsubsection{Reflection of Radio Signals}

If the detected signals were reflections from Starlink satellites, one potential origin could be solar radio emission. Solar activities are known to exhibit strong circular polarization at low frequencies \citep{2013ApJ...775...38S, 2019SoPh..294..106M}. Our observations were conducted in June and July 2024, during a period approaching the solar maximum\footnote{The Sun reached maximum phase in Oct 2024. \url{https://science.nasa.gov/science-research/heliophysics/nasa-noaa-sun-reaches-maximum-phase-in-11-year-solar-cycle/}} of the 11-year cycle, when increased solar activity was expected.

To test this hypothesis, we performed several nighttime observations, assuming that the satellites would be in Earth’s shadow and shielded from solar radio emission. Multiple detections were made at night. However, detailed calculations using TLE data revealed that all Starlink passes observed during our nighttime sessions were illuminated by the Sun, including the passes not detected. This was due to the timing of our observations near the summer solstice, when the Sun’s elevation at NenuFAR site (latitude +47 degrees) remained above approximately -23 degrees even at midnight. Given the satellites’ altitudes of several hundred kilometers above Earth’s surface, they remained illuminated by the Sun throughout day and night\footnote{This scenario resembles the "midnight sun" phenomenon at polar regions.}. Consequently, our nighttime detections could not conclusively rule out solar emission as a potential source.

Despite this, the broadband structure of the detected signals, particularly the distinct peak around 60 MHz, does not align with the characteristics of typical solar radio bursts, which generally exhibit frequency drifts or broad, continuous frequency ranges without distinct peaks \citep{2012SoPh..280..591Z,2024arXiv240500959W}. These discrepancies strongly suggest that the detected signals are not direct reflections of solar emission. However, preliminary results from LBA all-sky imaging with the LOFAR2.0 test stations (Bassa et al., in prep.) and early findings from NenuFAR exoplanetary observations (Zhang et al., in prep.) indicate that Starlink satellites appear fainter/undetectable when in Earth’s shadow, suggesting a possible link between the emissions and power generation via solar panels. Further dedicated studies are needed to confirm this correlation.

We also considered the possibility that the detected signals could be reflections of other bright celestial radio sources, particularly the Galactic Plane and strong radio sources such as Cygnus A and Cassiopeia A. These sources are among the brightest in the radio sky, emitting strongly across a broad range of frequencies, and their reflections have been proposed as a potential origin for low-frequency radio emissions associated with meteors \citep{2021JIMO...49..137D}. However, both the Galactic Plane and these bright sources are effectively unpolarized at low frequencies. Their synchrotron radiation, which is inherently dominated by linear polarization, undergoes significant depolarization due to Faraday effects \citep{2013AN....334..548B}.
Thus the significant circular polarization exhibited in Starlink detections is inconsistent with the expected polarization characteristics of these sources. Furthermore, the distinctive peak structure of the detected signals does not align with the emission spectra of either the Galactic Plane, Cygnus A, or Cassiopeia A, further disfavoring this hypothesis.

Another possibility we considered is that the detected emissions from Starlink satellites could be reflections of terrestrial signals, given that ground-based transmitters can produce strong signals within various frequency ranges. For example, television broadcasts in Europe are allocated near 60 MHz and are predominantly linearly polarized, as noted in the European Table of Frequency Allocations and Applications \footnote{\url{https://efis.cept.org}, accessed November 2024, CEPT/ERC Report 25}. However, the detected emissions from Starlink satellites exhibit significant circular polarization, which is uncommon for terrestrial transmitters at these frequencies. This disparity in polarization characteristics makes it unlikely that the observed signals originate as reflections of terrestrial broadcasts.

To further test the hypothesis that the detected emissions were reflections of terrestrial signals, we compared the detections from Starlink satellites with observations of other objects capable of reflecting ground-based signals: the International Space Station (ISS) and an Airbus A319 aircraft. Both the ISS and aircraft are significantly larger and closer to the Earth’s surface than Starlink satellites (The size and altitude information for STARLINK-31097, the ISS, and the Airbus A319 were obtained from online sources\footnote{\url{https://planet4589.org/astro/starsim/papers/StarGen2.pdf}}\footnote{\url{https://www.esa.int/Science_Exploration/Human_and_Robotic_Exploration/International_Space_Station/ISS_International_Space_Station}}\footnote{\url{https://aircraft.airbus.com/en/aircraft/a320-the-most-successful-aircraft-family-ever/a319neo}}.). According to the radar equation \citep{1980mgh..book.....S}:

\[
P_r = \frac{P_t G_t G_r \sigma \lambda^2}{(4 \pi)^3 R^4}
\]

where \(P_r\) is the received power, \(P_t\) is the transmitted power, \(G_t\) and \(G_r\) are the gains of the transmitting and receiving antennas, \(\sigma\) is the radar cross-section of the reflecting object, \(\lambda\) is the wavelength, and \(R\) is the distance between the transmitter and the reflecting object. This equation indicates that larger and closer objects should produce stronger reflected signals. 

If the Starlink detections were reflections of terrestrial signals, we would expect even stronger reflections from the ISS and aircraft due to their larger sizes and closer proximity. Specifically, reflections from the ISS are estimated to be 186.7 times stronger than those from Starlink satellites, while aircraft reflections are estimated to be approximately \(1.5 \times 10^8\) times brighter. \footnote{In practice, due to the relatively low altitude of aircraft, such reflections would require a transmitter positioned close to its flight path.}

\begin{figure*}
\centering
\includegraphics[width=\textwidth]{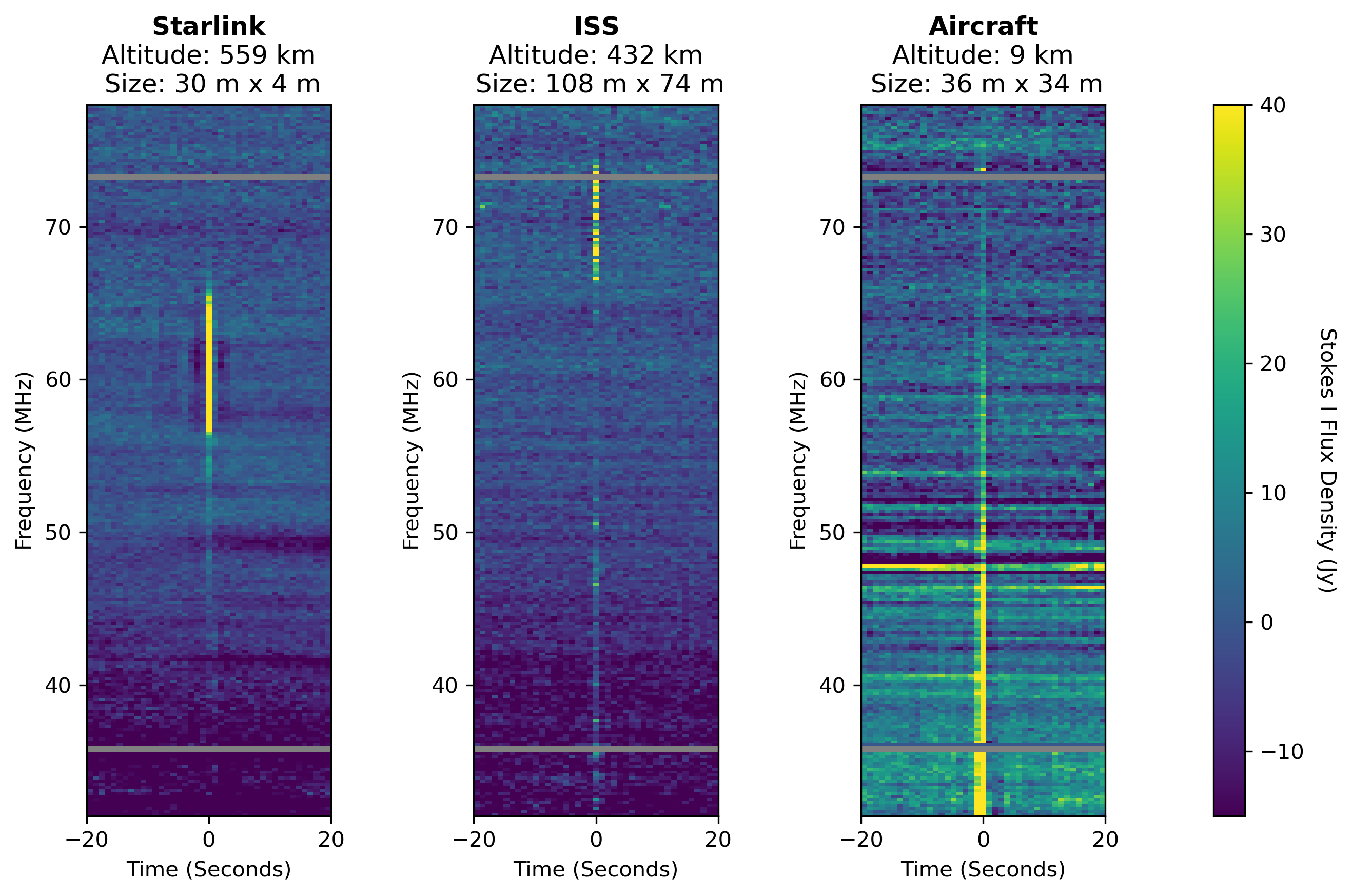}
\caption{Dynamic spectra of three objects observed with NenuFAR: STARLINK-31097, the International Space Station (ISS), and an Airbus A319 aircraft. The x-axis represents time, with 0 seconds corresponding to the time of the detection of the target. The altitudes of the objects, determined at the time of detection, and their sizes (length times width) are listed in each panel. Note that the noise level in the aircraft panel is higher compared to the other two panels, as the aircraft observation occurred during the daytime, while the Starlink and ISS observations were made at night. The grey-shaded regions in each panel indicate frequency channels known to be contaminated by RFI, which were masked during the analysis.}
\label{Fig:compare}
\end{figure*}

However, our observations did not detect any signals from the ISS or the aircraft at 60 MHz \footnote{Signals from the ISS and aircraft were detected at other frequencies. However, determining whether these emissions are intrinsic or reflections is beyond the scope of this paper.}, as shown in Figure~\ref{Fig:compare}. This result strongly suggests that the detected emissions from Starlink satellites are not simple reflections of terrestrial signals.

Finally, we evaluate the power requirements for a hypothetical transmitter under the assumption that the detected signals result from reflections. Our observations revealed broadband emission exceeding 200\,Jy from STARLINK-31034 over the 54--66\,MHz range at an altitude of 440 km. Given that each solar panel measures $15 \times 4\,\mathrm{m}$, we assume an ideal effective reflecting area of $120\,\mathrm{m}^2$ for two panels. Consequently, the radar equation yields two scenarios:

\begin{itemize}
    \item If the transmitter were astronomical, it would need to emit more than $4 \times 10^{12}$ Jy in this frequency range - a flux density far exceeding even the most active solar bursts \citep{2023BAAS...55c.429W}.
    \item If the transmitter were ground-based, its required power output would exceed 1.2 MW, which is substantially higher than that of most radio transmitters (e.g. TV broadcasters) operating in this frequency range.
\end{itemize}

Both cases present implausible conditions, making it unlikely that the detected signals originate from simple reflections.

\subsubsection{Unintended Intrinsic Emissions}

Given the exclusion of direct emissions in the allocated frequency bands and reflections of external signals, we next consider the possibility that the detected signals arise from unintended intrinsic emissions within the satellite systems themselves. Satellites, particularly those in complex constellations like Starlink, house sophisticated electronics, antennas, and power systems that may produce emissions outside their intended operational bands.

Unintended intrinsic emissions from satellites can originate from several sources, as outlined in studies of electromagnetic compatibility (EMC) in complex systems \citep{1992itec.book.....P}. These emissions often arise from the non-linear behavior of onboard electronic components, such as power supply circuits, oscillators, and high-speed digital processors. Switching noise from power supplies, particularly in systems using pulse-width modulation (PWM), can generate harmonics that propagate through the satellite’s structure. Similarly, high-speed digital circuits can produce transient electromagnetic fields due to rapid changes in current flow, leading to broad-spectrum emissions. Coupling between adjacent systems can exacerbate these effects by introducing additional resonances or feedback loops. The distinct peak observed around 60 MHz suggests the possibility of resonance effects within the satellite structure, potentially involving components such as solar panels acting as resonant elements at these frequencies. These complexities highlight the challenges of designing satellite systems that minimize such emissions while maintaining performance.

To explore whether the detected signals originate from unintended intrinsic emissions, we examined the detection rates and flux density levels across different generations of Starlink satellites. Starlink satellites can be broadly categorized into two primary versions: the earlier V0.9, V1.0 and V1.5 models, and the more recent V2-mini satellites. While both versions are designed to deliver broadband internet services, the V2-mini satellites introduce significant advancements. These include permission to operate at lower orbital altitudes, the addition of VHF beacon systems for telemetry, tracking, and command (TT\&C) during orbit-raising and emergencies, and improvements in autonomous collision avoidance systems\footnote{Details can be found in the FCC document DA-24-1193A1, available at \url{https://docs.fcc.gov/public/attachments/DA-24-1193A1.pdf} (accessed November 2024).}. By comparing detection rates and signal characteristics between these versions, we aim to assess how changes in satellite hardware might influence the properties of unintended emissions, providing valuable insights into their origins. A recent LOFAR study by \citet{2024A&A...689L..10B} highlights significant differences in emissions between the two Starlink generations, reinforcing the potential impact of hardware configurations on these signals.

\begin{figure}
\centering
\includegraphics[width=\columnwidth]{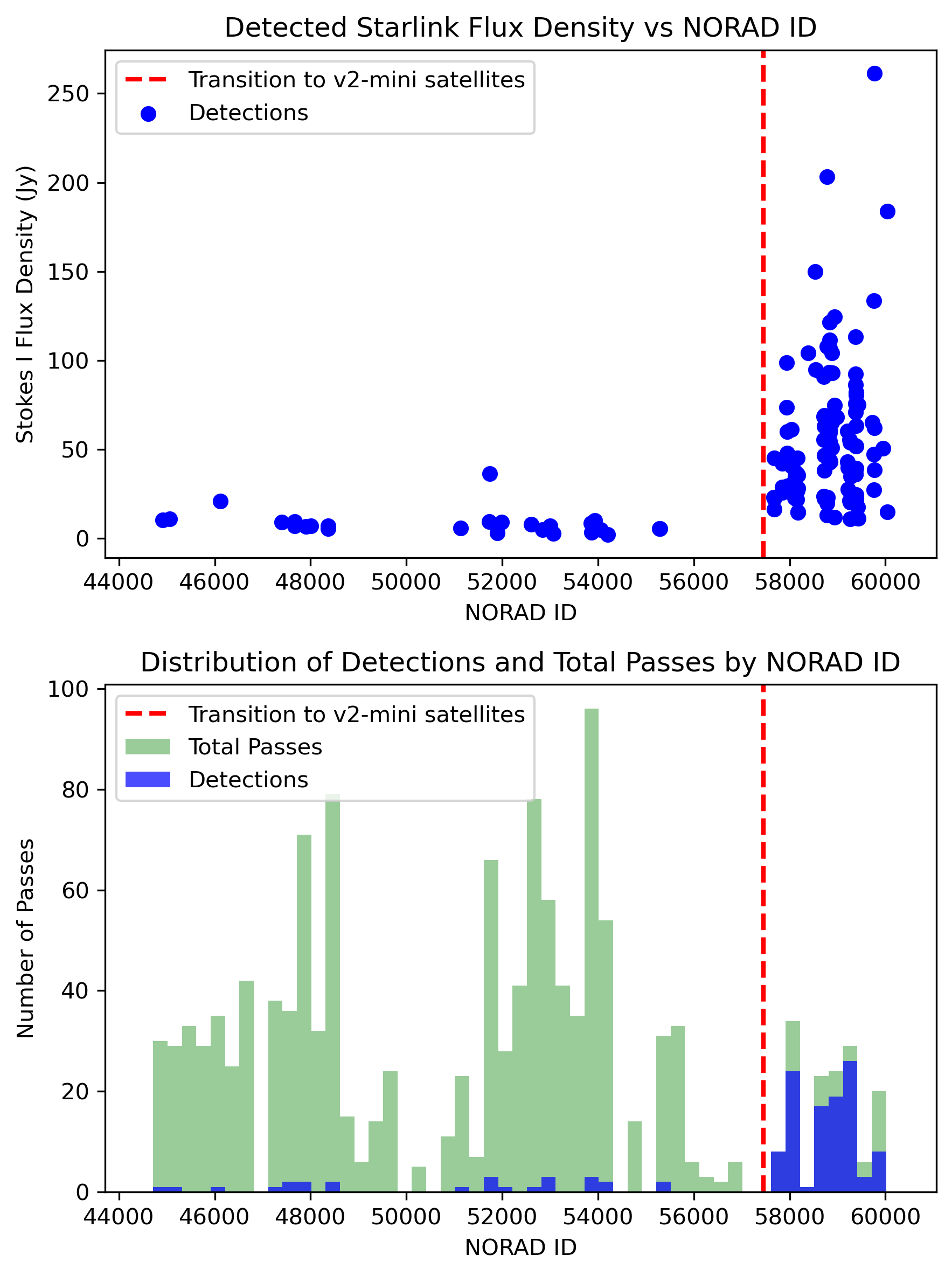}
\caption{Flux density and detection statistics of Starlink satellites by NORAD ID. The top panel shows the Stokes I flux density of detected satellites, integrated over the 54--66\,MHz range and 1-second intervals. The vertical dashed line at NORAD ID 57457 (corresponding to STARLINK-30165) indicates the transition to the newer V2-mini Starlink satellites. The bottom panel presents histograms of the number of detections (blue) and total satellite passes (green) as a function of NORAD ID, highlighting the increased detectability of the V2-mini satellites.}
\label{Fig:statistics}
\end{figure}

Based on these considerations, we analyzed the statistics of the older and newer Starlink satellites, as illustrated in Figure \ref{Fig:statistics}. The figure presents the flux density and detection rates of Starlink satellites as a function of their NORAD IDs—a unique identifier assigned to each artificial satellite. NORAD IDs are assigned sequentially, with earlier IDs corresponding to satellites launched earlier in time.

Our observations reveal a striking contrast between the two versions of Starlink satellites. Nearly all passes of the newer V2-mini satellites were detected within NenuFAR’s observational FoV, with flux density levels frequently surpassing those of the older versions, reaching a few hundreds of Jy. In comparison, the older satellites yielded detectable signals during only a few percents of total passes, with all detections exhibiting flux density levels below 50 Jy. 

Given that the reflective properties of both satellite generations should be similar, the stark difference in detection rates and flux density levels argues against the reflection hypothesis as the primary cause of the observed signals. Instead, the upgraded systems on the V2-mini satellites may be responsible for their increased detectability. These emissions are likely unintended and incidental, arising as byproducts of the new hardware configurations or enhanced power systems.

\subsection{Impacts on Radio Astronomy}

The increasing presence of Starlink satellites in orbit poses significant challenges to various aspects of radio astronomy. These challenges include technical issues during data calibration and disruptions in specific scientific fields. Addressing them requires thoughtful mitigation strategies.

\subsubsection{Technical Challenges: Calibration and Near-Field Effects}
The transient and bright nature of Starlink satellites poses significant challenges to calibration in low-frequency interferometry, particularly self-calibration. These challenges become critical when the solution intervals of self-calibration are at the scale of a minute or less—comparable to the time it takes for a satellite to traverse the FoV. Self-calibration relies on accurate sky models that map the distribution of flux density from known celestial sources. However, since the transient flux density from Starlink satellites is difficult to incorporate into sky models, their presence introduces unmodeled signals into the observations, leading to errors in the reconstructed flux densities of celestial sources.

These challenges are further amplified in advanced calibration methods, such as polarimetric calibration and direction-dependent calibration, due to the limited number of sources included in the relevant sky models. At GHz frequencies, polarized sources—such as radio galaxies and pulsars—are prominent, but at frequencies below 100 MHz, most of these sources are depolarized due to Faraday rotation across the observing bandwidth and spatial variations in polarization angle \citep{2017PASA...34...40L}, leaving very few celestial sources intrinsically polarized at these frequencies. As a result, it is often assumed that all sources in the sky model are unpolarized to simplify the calibration process. However, Starlink satellites exhibit significant polarization, making them dominant contributors to the observed polarized sky. In a full-Stokes calibration where Q, U and V components are fitted, the presence of polarized flux density from these satellites violates the unpolarized assumption, causing genuinely unpolarized sources in the sky model to appear artificially polarized, thereby generating spurious artifacts in the data.

Direction-dependent calibration introduces additional complications. In this method, the FoV is divided into smaller regions, each assigned its own calibration solution to account for spatially varying instrumental and ionospheric effects \citep{2011MNRAS.414.1656K, 2021A&A...648A...1T}. Although a single satellite pass may not dominate the total flux density across the entire FoV, its localized brightness can significantly affect the calibration of individual facets. This localized distortion can degrade the quality of calibration solutions across the FoV, reducing the overall fidelity of the reconstructed images.

In addition to these calibration challenges, another important consideration is the near-field nature of Starlink satellites. While standard interferometric imaging assumes sources to be in the far field, satellites in LEO can violate this assumption, especially for arrays with long baselines. In the near field, wavefront curvature across the array becomes non-negligible, potentially leading to decorrelation and reduced calibration accuracy if not accounted for. Recent work has demonstrated near-field aperture synthesis techniques that restore coherence for LEO satellites \citep{2023PASA...40...56P}, as well as methods to correct for near-field effects and estimate emitter altitudes using interferometric data \citep{2025PASA...42...10D}. These methods may become increasingly important as satellite constellations grow.

\subsubsection{Scientific Challenges: Transient and Non-Transient Studies}
Even with accurate calibration, the presence of Starlink satellites still impacts data products across various research fields.

For transient studies, including pulsars, fast radio bursts (FRBs), long-period transients \citep{2022Natur.601..526H,2023Natur.619..487H}, exoplanets, and flare stars, the bright bursts introduced by Starlink satellites into dynamic spectra complicate the identification of genuine astrophysical signals. These bursts require additional verification steps for candidate detections. One straightforward approach is imaging: Starlink satellites move rapidly across the sky, with angular velocities of 1–2 degrees per second near the zenith\footnote{Calculated using relevant TLE data.}. By creating images with integration times of a few seconds, these satellites leave distinct trails that clearly indicate contamination, which can be removed via multi-dimention masking \citep{2024A&A...681A..71G}. Another method involves dispersion tests: many astrophysical transients exhibit nonzero dispersion measures (DM) due to their propagation through the interstellar medium, whereas satellite-induced bursts remain at DM = 0. Dispersion tests provide a reliable way to differentiate genuine astrophysical sources from satellite-induced signals. In this case, a compromise must be chosen from the point of view of ensuring a sufficiently wide frequency band, frequency and time resolution, and the proposed new criterion for distinguishing from satellite signals - the spatial criterion.

Non-transient studies, such as surveys, cosmic magnetism, and 21-cm cosmology research \citep{2024A&A...681A..71G}, face significant challenges from the presence of Starlink satellites. While emissions from individual satellites may average out in datasets stacked over multiple hours, the cumulative effect of numerous satellite trails\footnote{For example, the detection rate of Starlink trails is approximately one trail per 15 minutes of observation for NenuFAR, as discussed in Section \ref{sec:results}.} introduces additional noise, which can be referred to as the “Satellite Meterwave Foreground.” This added noise elevates the overall noise level in the final data products, potentially masking faint astrophysical signals and complicating foreground removal techniques.

\subsubsection{Potential Mitigation Strategies}
Addressing the impact of Starlink satellites on radio astronomy requires targeted mitigation strategies. A common approach involves multi-dimensional masking techniques \citep{2024A&A...681A..71G}, which filter transient RFI in the image domain. However, since satellite trajectories are predictable, a simpler approach is to cross-match observational data with predicted satellite passes and flag contaminated time segments. Both methods can effectively reduce satellite-induced noise but come at the cost of decreasing the amount of usable data. More advanced solutions are also being explored, such as the TABASCAL project \citep{2023MNRAS.524.3231F}, which proposes adaptive calibration techniques capable of modeling moving satellite signals, but practical implementation remains extremely challenging.

As of July 2024, there are approximately 1,000 V2-mini satellites in orbit\footnote{The list of satellites is derived from Space-Track, \url{https://www.space-track.org/}.}, which dominate the Starlink-related RFI below 100 MHz. For a wide-FoV array such as NenuFAR, excluding time segments when a V2-mini satellite is within the imaging FoV results in a loss of around 1.5\% of observational data\footnote{This estimate is based on the typical satellite passage duration of approximately 15 seconds per 15-minute observation window.} — an acceptable compromise for many studies. However, SpaceX plans to launch up to 29,988 additional V2 satellites\footnote{Details are provided in the FCC document DA-24-1193A1, available at \url{https://docs.fcc.gov/public/attachments/DA-24-1193A1.pdf}.}. If similar emissions persist in future launches, the fraction of contaminated data could rise to approximately 45\%. Such a significant data loss would severely hinder long-duration studies, posing serious challenges to research across multiple fields of low-frequency radio astronomy, including experiments targeting the faint 21-cm signal from the early Universe \citep{2020MNRAS.498..265W}. Moreover, this estimate only accounts for Starlink, while additional planned mega-constellations, such as Amazon’s Project Kuiper\footnote{\url{https://www.aboutamazon.com/what-we-do/devices-services/project-kuiper}}, could further exacerbate the impact.

Collaboration between the astronomical community and satellite operators offers us a necessary pathway to mitigate these challenges. Examples of such efforts include SpaceX's "Darksat," which reduced optical brightness by approximately 0.77 magnitudes \citep{2020A&A...637L...1T}, and the Green Bank Telescope (GBT) collaboration, where SpaceX implemented a boresight avoidance strategy to significantly reduce downlink RFI \citep{2024ApJ...971L..49N}. These initiatives highlight the potential of coordinated solutions, such as incorporating RFI mitigation into satellite design, developing predictive contamination models for satellites, and implementing operational protocols. Sustained collaboration is essential to ensure the future sustainability of low-frequency radio astronomy in the era of satellite mega-constellations.

\section{Conclusions}
\label{sec:conclusion}

We report the detection and analysis of broadband radio emissions from Starlink satellites below 100 MHz using NenuFAR. Over 40 hours of observations, we identified 134 satellite passes, with flux densities ranging from 2 Jy to 261 Jy. The emission exhibited a broadband spectrum with distinctive features, including a “double-peak” structure between 54–66 MHz and significant circular polarization. Our analysis indicates that the detected signals are likely unintended intrinsic emissions from the satellites, associated with hardware configurations or onboard electronics, particularly in the newer V2-mini satellites.

These detections present substantial challenges to radio astronomy. The transient and polarized nature of the emissions disrupts calibration processes, particularly polarimetric and direction-dependent calibration, and complicates transient searches by introducing false signals. Non-transient studies face increased noise levels from cumulative satellite trails, which could obscure faint astrophysical signals and reduce observational sensitivity. These issues are expected to worsen significantly with the planned expansion of satellite mega-constellations.

Mitigation strategies, including data flagging and exclusion of contaminated time segments, offer partial relief but are insufficient in the long term. As the number of satellites increases, collaborative efforts between astronomers and satellite operators will be crucial. Collaborative efforts between the astronomical community and satellite operators have shown promise, particularly in addressing challenges in optical and radio astronomy. Future initiatives should focus on improving satellite designs, refining operations, and developing tools to predict and mitigate satellite contamination in radio observations.

\begin{acknowledgements}
      The authors acknowledge funding from the ERC under the European Union’s Horizon 2020 research and innovation programme (grant agreement no. 101020459 - Exoradio).

      This work is based on data obtained using the NenuFAR radiotelescope. NenuFAR has benefited from the funding from CNRS/INSU, Observatoire de Paris, Observatoire Radioastronomique de Nançay, Université d’Orléans, Région Centre-Val de Loire, DIM-ACAV -ACAV+ \& -Origines de la Région Ile de France, and Agence Nationale de la Recherche. We acknowledge the collective work from the NenuFAR-France collaboration for making NenuFAR operational, and the Nançay Data Center resources used for data reduction and storage.

      X.Z. acknowledges the support of a postdoctoral fellowship at CSIRO, Australia, where the low-frequency polarimetry techniques developed during that time contributed to this work.

      X.Z. thanks L. Lamy for the insightful discussions on the potential origins of the detected signals.

      J.D.T. was supported for this work by the TESS Guest Investigator Program G06165 and by NASA through the NASA Hubble Fellowship grant $\#$HST-HF2-51495.001-A awarded by the Space Telescope Science Institute, which is operated by the Association of Universities for Research in Astronomy, Incorporated, under NASA contract NAS5-26555.

      L.V.E.K. acknowledges the financial support from the European Research Council (ERC) under the European Union’s Horizon 2020 research and innovation programme (Grant agreement No. 884760, “CoDEX”)

      The authors acknowledge the use of AI-assisted copy editing (ChatGPT) to improve the text of the manuscript.
\end{acknowledgements}

\bibliographystyle{aa} 
\bibliography{aanda} 

\begin{thebibliography}{59}
\expandafter\ifx\csname natexlab\endcsname\relax\def\natexlab#1{#1}\fi

\bibitem[{{Bassa} {et~al.}(2024){Bassa}, {Di Vruno}, {Winkel}, {J{\'o}zsa}, {Brentjens}, \& {Zhang}}]{2024A&A...689L..10B}
{Bassa}, C.~G., {Di Vruno}, F., {Winkel}, B., {et~al.} 2024, \aap, 689, L10

\bibitem[{{Beck} {et~al.}(2013){Beck}, {Anderson}, {Heald}, {Horneffer}, {Iacobelli}, {K{\"o}hler}, {Mulcahy}, {Pizzo}, {Scaife}, {Wucknitz}, \& {LOFAR Magnetism Key Science Project Team}}]{2013AN....334..548B}
{Beck}, R., {Anderson}, J., {Heald}, G., {et~al.} 2013, Astronomische Nachrichten, 334, 548

\bibitem[{{Brentjens} \& {de Bruyn}(2005)}]{2005A&A...441.1217B}
{Brentjens}, M.~A. \& {de Bruyn}, A.~G. 2005, \aap, 441, 1217

\bibitem[{{Di Vruno} {et~al.}(2023){Di Vruno}, {Winkel}, {Bassa}, {J{\'o}zsa}, {Brentjens}, {Jessner}, \& {Garrington}}]{2023A&A...676A..75D}
{Di Vruno}, F., {Winkel}, B., {Bassa}, C.~G., {et~al.} 2023, \aap, 676, A75

\bibitem[{{Dijkema} {et~al.}(2021){Dijkema}, {Bassa}, {Kuiack}, {Jenniskens}, {Johannink}, {Bettonvil}, {Wijers}, \& {Fallows}}]{2021JIMO...49..137D}
{Dijkema}, T.~J., {Bassa}, C., {Kuiack}, M., {et~al.} 2021, WGN, Journal of the International Meteor Organization, 49, 137

\bibitem[{{Ducharme} \& {Pober}(2025)}]{2025PASA...42...10D}
{Ducharme}, J.~M. \& {Pober}, J.~C. 2025, \pasa, 42, e010

\bibitem[{{Finlay} {et~al.}(2023){Finlay}, {Bassett}, {Kunz}, \& {Oozeer}}]{2023MNRAS.524.3231F}
{Finlay}, C., {Bassett}, B.~A., {Kunz}, M., \& {Oozeer}, N. 2023, \mnras, 524, 3231

\bibitem[{{Gallozzi} {et~al.}(2020){Gallozzi}, {Scardia}, \& {Maris}}]{2020arXiv200110952G}
{Gallozzi}, S., {Scardia}, M., \& {Maris}, M. 2020, arXiv e-prints, arXiv:2001.10952

\bibitem[{{Gehlot} {et~al.}(2024){Gehlot}, {Koopmans}, {Brackenhoff}, {Ceccotti}, {Ghosh}, {H{\"o}fer}, {Mertens}, {Mevius}, {Munshi}, {Offringa}, {Pandey}, {Rowlinson}, {Shulevski}, {Wijers}, {Yatawatta}, \& {Zaroubi}}]{2024A&A...681A..71G}
{Gehlot}, B.~K., {Koopmans}, L.~V.~E., {Brackenhoff}, S.~A., {et~al.} 2024, \aap, 681, A71

\bibitem[{Grie{\ss}meier {et~al.}(2021)Grie{\ss}meier, Brionne, Kondratiev, Bilous, McKee, Zarka, Viou, {et~al.}}]{bondonneau2021pulsars}
Grie{\ss}meier, J.-M., Brionne, M., Kondratiev, V., {et~al.} 2021, Astronomy \& Astrophysics, 652, A34

\bibitem[{Grigg {et~al.}(2025)Grigg, Tingay, Prabu, Sokolowski, \& Indermuehle}]{grigg2025enhanced}
Grigg, D., Tingay, S., Prabu, S., Sokolowski, M., \& Indermuehle, B. 2025, Publications of the Astronomical Society of Australia, 42, e015

\bibitem[{{Grigg} {et~al.}(2023){Grigg}, {Tingay}, {Sokolowski}, {Wayth}, {Indermuehle}, \& {Prabu}}]{2023A&A...678L...6G}
{Grigg}, D., {Tingay}, S.~J., {Sokolowski}, M., {et~al.} 2023, \aap, 678, L6

\bibitem[{{Hainaut} \& {Williams}(2020)}]{2020A&A...636A.121H}
{Hainaut}, O.~R. \& {Williams}, A.~P. 2020, \aap, 636, A121

\bibitem[{{Hurley-Walker} {et~al.}(2023){Hurley-Walker}, {Rea}, {McSweeney}, {Meyers}, {Lenc}, {Heywood}, {Hyman}, {Men}, {Clarke}, {Coti Zelati}, {Price}, {Horv{\'a}th}, {Galvin}, {Anderson}, {Bahramian}, {Barr}, {Bhat}, {Caleb}, {Dall'Ora}, {de Martino}, {Giacintucci}, {Morgan}, {Rajwade}, {Stappers}, \& {Williams}}]{2023Natur.619..487H}
{Hurley-Walker}, N., {Rea}, N., {McSweeney}, S.~J., {et~al.} 2023, \nat, 619, 487

\bibitem[{{Hurley-Walker} {et~al.}(2022){Hurley-Walker}, {Zhang}, {Bahramian}, {McSweeney}, {O'Doherty}, {Hancock}, {Morgan}, {Anderson}, {Heald}, \& {Galvin}}]{2022Natur.601..526H}
{Hurley-Walker}, N., {Zhang}, X., {Bahramian}, A., {et~al.} 2022, \nat, 601, 526

\bibitem[{{Kazemi} {et~al.}(2011){Kazemi}, {Yatawatta}, {Zaroubi}, {Lampropoulos}, {de Bruyn}, {Koopmans}, \& {Noordam}}]{2011MNRAS.414.1656K}
{Kazemi}, S., {Yatawatta}, S., {Zaroubi}, S., {et~al.} 2011, \mnras, 414, 1656

\bibitem[{{Kondratiev} {et~al.}(2016){Kondratiev}, {Verbiest}, {Hessels}, {Bilous}, {Stappers}, {Kramer}, {Keane}, {Noutsos}, {Os{\l}owski}, {Breton}, {Hassall}, {Alexov}, {Cooper}, {Falcke}, {Grie{\ss}meier}, {Karastergiou}, {Kuniyoshi}, {Pilia}, {Sobey}, {ter Veen}, {van Leeuwen}, {Weltevrede}, {Bell}, {Broderick}, {Corbel}, {Eisl{\"o}ffel}, {Markoff}, {Rowlinson}, {Swinbank}, {Wijers}, {Wijnands}, \& {Zarka}}]{2016A&A...585A.128K}
{Kondratiev}, V.~I., {Verbiest}, J.~P.~W., {Hessels}, J.~W.~T., {et~al.} 2016, \aap, 585, A128

\bibitem[{{Koopmans} {et~al.}(2015){Koopmans}, {Pritchard}, {Mellema}, {Aguirre}, {Ahn}, {Barkana}, {van Bemmel}, {Bernardi}, {Bonaldi}, {Briggs}, {de Bruyn}, {Chang}, {Chapman}, {Chen}, {Ciardi}, {Dayal}, {Ferrara}, {Fialkov}, {Fiore}, {Ichiki}, {Illiev}, {Inoue}, {Jelic}, {Jones}, {Lazio}, {Maio}, {Majumdar}, {Mack}, {Mesinger}, {Morales}, {Parsons}, {Pen}, {Santos}, {Schneider}, {Semelin}, {de Souza}, {Subrahmanyan}, {Takeuchi}, {Vedantham}, {Wagg}, {Webster}, {Wyithe}, {Datta}, \& {Trott}}]{2015aska.confE...1K}
{Koopmans}, L., {Pritchard}, J., {Mellema}, G., {et~al.} 2015, in Advancing Astrophysics with the Square Kilometre Array (AASKA14), 1

\bibitem[{{Kruk} {et~al.}(2023){Kruk}, {Garc{\'\i}a-Mart{\'\i}n}, {Popescu}, {Aussel}, {Dillmann}, {Perks}, {Lund}, {Mer{\'\i}n}, {Thomson}, {Karadag}, \& {McCaughrean}}]{2023NatAs...7..262K}
{Kruk}, S., {Garc{\'\i}a-Mart{\'\i}n}, P., {Popescu}, M., {et~al.} 2023, Nature Astronomy, 7, 262

\bibitem[{Labate {et~al.}(2022)Labate, Waterson, Alachkar, Hendre, Lewis, Bartolini, \& Dewdney}]{labate2022highlights}
Labate, M.~G., Waterson, M., Alachkar, B., {et~al.} 2022, Journal of Astronomical Telescopes, Instruments, and Systems, 8, 011024

\bibitem[{{Lazio} {et~al.}(2004){Lazio}, {Farrell}, {Dietrick}, {Greenlees}, {Hogan}, {Jones}, \& {Hennig}}]{2004ApJ...612..511L}
{Lazio}, T.~Joseph, W., {Farrell}, W.~M., {Dietrick}, J., {et~al.} 2004, \apj, 612, 511

\bibitem[{{Lenc} {et~al.}(2017){Lenc}, {Anderson}, {Barry}, {Bowman}, {Cairns}, {Farnes}, {Gaensler}, {Heald}, {Johnston-Hollitt}, {Kaplan}, {Lynch}, {McCauley}, {Mitchell}, {Morgan}, {Morales}, {Murphy}, {Offringa}, {Ord}, {Pindor}, {Riseley}, {Sadler}, {Sobey}, {Sokolowski}, {Sullivan}, {O'Sullivan}, {Sun}, {Tremblay}, {Trott}, \& {Wayth}}]{2017PASA...34...40L}
{Lenc}, E., {Anderson}, C.~S., {Barry}, N., {et~al.} 2017, \pasa, 34, e040

\bibitem[{{Loh} \& the NenuFAR~team(2020)}]{alan_loh_2020_4279405}
{Loh}, A. \& the NenuFAR~team. 2020, nenupy: a Python package for the low-frequency radio telescope NenuFAR

\bibitem[{{McCauley} {et~al.}(2019){McCauley}, {Cairns}, {White}, {Mondal}, {Lenc}, {Morgan}, \& {Oberoi}}]{2019SoPh..294..106M}
{McCauley}, P.~I., {Cairns}, I.~H., {White}, S.~M., {et~al.} 2019, \solphys, 294, 106

\bibitem[{{McDowell}(2020)}]{2020ApJ...892L..36M}
{McDowell}, J.~C. 2020, \apjl, 892, L36

\bibitem[{{Mellema} {et~al.}(2013){Mellema}, {Koopmans}, {Abdalla}, {Bernardi}, {Ciardi}, {Daiboo}, {de Bruyn}, {Datta}, {Falcke}, {Ferrara}, {Iliev}, {Iocco}, {Jeli{\'c}}, {Jensen}, {Joseph}, {Labroupoulos}, {Meiksin}, {Mesinger}, {Offringa}, {Pandey}, {Pritchard}, {Santos}, {Schwarz}, {Semelin}, {Vedantham}, {Yatawatta}, \& {Zaroubi}}]{2013ExA....36..235M}
{Mellema}, G., {Koopmans}, L. V.~E., {Abdalla}, F.~A., {et~al.} 2013, Experimental Astronomy, 36, 235

\bibitem[{Mevius(2018)}]{mevius2018rmextract}
Mevius, M. 2018, Astrophysics Source Code Library, ascl

\bibitem[{Munshi {et~al.}(2024)Munshi, Mertens, Koopmans, Offringa, Semelin, Aubert, Barkana, Bracco, Brackenhoff, Cecconi, {et~al.}}]{munshi2024first}
Munshi, S., Mertens, F., Koopmans, L., {et~al.} 2024, Astronomy \& Astrophysics, 681, A62

\bibitem[{{Munshi} {et~al.}(2024){Munshi}, {Mertens}, {Koopmans}, {Offringa}, {Semelin}, {Aubert}, {Barkana}, {Bracco}, {Brackenhoff}, {Cecconi}, {Ceccotti}, {Corbel}, {Fialkov}, {Gehlot}, {Ghara}, {Girard}, {Grie{\ss}meier}, {H{\"o}fer}, {Hothi}, {M{\'e}riot}, {Mevius}, {Ocvirk}, {Shaw}, {Theureau}, {Yatawatta}, {Zarka}, \& {Zaroubi}}]{2024A&A...681A..62M}
{Munshi}, S., {Mertens}, F.~G., {Koopmans}, L.~V.~E., {et~al.} 2024, \aap, 681, A62

\bibitem[{{Nhan} {et~al.}(2024){Nhan}, {De Pree}, {Iverson}, {Gregory}, {Dueri}, {Beasley}, \& {Schepis}}]{2024ApJ...971L..49N}
{Nhan}, B.~D., {De Pree}, C.~G., {Iverson}, M., {et~al.} 2024, \apjl, 971, L49

\bibitem[{Offringa {et~al.}(2014)Offringa, McKinley, Hurley-Walker, {et~al.}}]{offringa-wsclean-2014}
Offringa, A.~R., McKinley, B., Hurley-Walker, {et~al.} 2014, MNRAS, 444, 606

\bibitem[{{Offringa} {et~al.}(2012){Offringa}, {van de Gronde}, \& {Roerdink}}]{2012A&A...539A..95O}
{Offringa}, A.~R., {van de Gronde}, J.~J., \& {Roerdink}, J.~B.~T.~M. 2012, \aap, 539, A95

\bibitem[{{Paul}(1992)}]{1992itec.book.....P}
{Paul}, C.~R. 1992, {Introduction to electromagnetic compatibility}

\bibitem[{{Prabu} {et~al.}(2020){Prabu}, {Hancock}, {Zhang}, \& {Tingay}}]{2020PASA...37...52P}
{Prabu}, S., {Hancock}, P., {Zhang}, X., \& {Tingay}, S.~J. 2020, \pasa, 37, e052

\bibitem[{{Prabu} {et~al.}(2023){Prabu}, {Tingay}, \& {Williams}}]{2023PASA...40...56P}
{Prabu}, S., {Tingay}, S.~J., \& {Williams}, A. 2023, \pasa, 40, e056

\bibitem[{{Pritchard} \& {Loeb}(2012)}]{2012RPPh...75h6901P}
{Pritchard}, J.~R. \& {Loeb}, A. 2012, Reports on Progress in Physics, 75, 086901

\bibitem[{{Sasikumar Raja} \& {Ramesh}(2013)}]{2013ApJ...775...38S}
{Sasikumar Raja}, K. \& {Ramesh}, R. 2013, \apj, 775, 38

\bibitem[{{Skolnik}(1980)}]{1980mgh..book.....S}
{Skolnik}, M.~I. 1980, {Introduction to radar systems /2nd edition/}

\bibitem[{{Stappers} {et~al.}(2011){Stappers}, {Hessels}, {Alexov}, {Anderson}, {Coenen}, {Hassall}, {Karastergiou}, {Kondratiev}, {Kramer}, {van Leeuwen}, {Mol}, {Noutsos}, {Romein}, {Weltevrede}, {Fender}, {Wijers}, {B{\"a}hren}, {Bell}, {Broderick}, {Daw}, {Dhillon}, {Eisl{\"o}ffel}, {Falcke}, {Griessmeier}, {Law}, {Markoff}, {Miller-Jones}, {Scheers}, {Spreeuw}, {Swinbank}, {Ter Veen}, {Wise}, {Wucknitz}, {Zarka}, {Anderson}, {Asgekar}, {Avruch}, {Beck}, {Bennema}, {Bentum}, {Best}, {Bregman}, {Brentjens}, {van de Brink}, {Broekema}, {Brouw}, {Br{\"u}ggen}, {de Bruyn}, {Butcher}, {Ciardi}, {Conway}, {Dettmar}, {van Duin}, {van Enst}, {Garrett}, {Gerbers}, {Grit}, {Gunst}, {van Haarlem}, {Hamaker}, {Heald}, {Hoeft}, {Holties}, {Horneffer}, {Koopmans}, {Kuper}, {Loose}, {Maat}, {McKay-Bukowski}, {McKean}, {Miley}, {Morganti}, {Nijboer}, {Noordam}, {Norden}, {Olofsson}, {Pandey-Pommier}, {Polatidis}, {Reich}, {R{\"o}ttgering}, {Schoenmakers}, {Sluman}, {Smirnov}, {Steinmetz}, {Sterks}, {Tagger}, {Tang},
  {Vermeulen}, {Vermaas}, {Vogt}, {de Vos}, {Wijnholds}, {Yatawatta}, \& {Zensus}}]{2011A&A...530A..80S}
{Stappers}, B.~W., {Hessels}, J.~W.~T., {Alexov}, A., {et~al.} 2011, \aap, 530, A80

\bibitem[{{Tasse}(2023)}]{2023ascl.soft05005T}
{Tasse}, C. 2023, {killMS: Direction-dependent radio interferometric calibration package}, Astrophysics Source Code Library, record ascl:2305.005

\bibitem[{{Tasse} {et~al.}(2025){Tasse}, {Hardcastle}, {Zarka}, {Loh}, {Shimwell}, {Callingham}, \& {Hugo}}]{Tasse2025}
{Tasse}, C., {Hardcastle}, M., {Zarka}, P., {et~al.} 2025, \aap, submitted

\bibitem[{{Tasse} {et~al.}(2018){Tasse}, {Hugo}, {Mirmont}, {Smirnov}, {Atemkeng}, {Bester}, {Hardcastle}, {Lakhoo}, {Perkins}, \& {Shimwell}}]{2018A&A...611A..87T}
{Tasse}, C., {Hugo}, B., {Mirmont}, M., {et~al.} 2018, \aap, 611, A87

\bibitem[{{Tasse} {et~al.}(2021){Tasse}, {Shimwell}, {Hardcastle}, {O'Sullivan}, {van Weeren}, {Best}, {Bester}, {Hugo}, {Smirnov}, {Sabater}, {Calistro-Rivera}, {de Gasperin}, {Morabito}, {R{\"o}ttgering}, {Williams}, {Bonato}, {Bondi}, {Botteon}, {Br{\"u}ggen}, {Brunetti}, {Chy{\.z}y}, {Garrett}, {G{\"u}rkan}, {Jarvis}, {Kondapally}, {Mandal}, {Prandoni}, {Repetti}, {Retana-Montenegro}, {Schwarz}, {Shulevski}, \& {Wiaux}}]{2021A&A...648A...1T}
{Tasse}, C., {Shimwell}, T., {Hardcastle}, M.~J., {et~al.} 2021, \aap, 648, A1

\bibitem[{Toh {et~al.}(2003)Toh, Cahill, \& Fusco}]{toh2003understanding}
Toh, B.~Y., Cahill, R., \& Fusco, V.~F. 2003, IEEE Transactions on Education, 46, 313

\bibitem[{{Tregloan-Reed} {et~al.}(2020){Tregloan-Reed}, {Otarola}, {Ortiz}, {Molina}, {Anais}, {Gonz{\'a}lez}, {Colque}, \& {Unda-Sanzana}}]{2020A&A...637L...1T}
{Tregloan-Reed}, J., {Otarola}, A., {Ortiz}, E., {et~al.} 2020, \aap, 637, L1

\bibitem[{Vallado \& Cefola(2012)}]{vallado2012two}
Vallado, D.~A. \& Cefola, P.~J. 2012, in 63rd International Astronautical Congress, Naples, Italy, 1--14

\bibitem[{{van Diepen} {et~al.}(2018){van Diepen}, {Dijkema}, \& {Offringa}}]{2018ascl.soft04003V}
{van Diepen}, G., {Dijkema}, T.~J., \& {Offringa}, A. 2018, {DPPP: Default Pre-Processing Pipeline}, Astrophysics Source Code Library, record ascl:1804.003

\bibitem[{{van Haarlem} {et~al.}(2013){van Haarlem}, {Wise}, {Gunst}, {Heald}, {McKean}, {Hessels}, {de Bruyn}, {Nijboer}, {Swinbank}, {Fallows}, {Brentjens}, {Nelles}, {Beck}, {Falcke}, {Fender}, {H{\"o}randel}, {Koopmans}, {Mann}, {Miley}, {R{\"o}ttgering}, {Stappers}, {Wijers}, {Zaroubi}, {van den Akker}, {Alexov}, {Anderson}, {Anderson}, {van Ardenne}, {Arts}, {Asgekar}, {Avruch}, {Batejat}, {B{\"a}hren}, {Bell}, {Bell}, {van Bemmel}, {Bennema}, {Bentum}, {Bernardi}, {Best}, {B{\^\i}rzan}, {Bonafede}, {Boonstra}, {Braun}, {Bregman}, {Breitling}, {van de Brink}, {Broderick}, {Broekema}, {Brouw}, {Br{\"u}ggen}, {Butcher}, {van Cappellen}, {Ciardi}, {Coenen}, {Conway}, {Coolen}, {Corstanje}, {Damstra}, {Davies}, {Deller}, {Dettmar}, {van Diepen}, {Dijkstra}, {Donker}, {Doorduin}, {Dromer}, {Drost}, {van Duin}, {Eisl{\"o}ffel}, {van Enst}, {Ferrari}, {Frieswijk}, {Gankema}, {Garrett}, {de Gasperin}, {Gerbers}, {de Geus}, {Grie{\ss}meier}, {Grit}, {Gruppen}, {Hamaker}, {Hassall}, {Hoeft}, {Holties},
  {Horneffer}, {van der Horst}, {van Houwelingen}, {Huijgen}, {Iacobelli}, {Intema}, {Jackson}, {Jelic}, {de Jong}, {Juette}, {Kant}, {Karastergiou}, {Koers}, {Kollen}, {Kondratiev}, {Kooistra}, {Koopman}, {Koster}, {Kuniyoshi}, {Kramer}, {Kuper}, {Lambropoulos}, {Law}, {van Leeuwen}, {Lemaitre}, {Loose}, {Maat}, {Macario}, {Markoff}, {Masters}, {McFadden}, {McKay-Bukowski}, {Meijering}, {Meulman}, {Mevius}, {Middelberg}, {Millenaar}, {Miller-Jones}, {Mohan}, {Mol}, {Morawietz}, {Morganti}, {Mulcahy}, {Mulder}, {Munk}, {Nieuwenhuis}, {van Nieuwpoort}, {Noordam}, {Norden}, {Noutsos}, {Offringa}, {Olofsson}, {Omar}, {Orr{\'u}}, {Overeem}, {Paas}, {Pandey-Pommier}, {Pandey}, {Pizzo}, {Polatidis}, {Rafferty}, {Rawlings}, {Reich}, {de Reijer}, {Reitsma}, {Renting}, {Riemers}, {Rol}, {Romein}, {Roosjen}, {Ruiter}, {Scaife}, {van der Schaaf}, {Scheers}, {Schellart}, {Schoenmakers}, {Schoonderbeek}, {Serylak}, {Shulevski}, {Sluman}, {Smirnov}, {Sobey}, {Spreeuw}, {Steinmetz}, {Sterks}, {Stiepel}, {Stuurwold},
  {Tagger}, {Tang}, {Tasse}, {Thomas}, {Thoudam}, {Toribio}, {van der Tol}, {Usov}, {van Veelen}, {van der Veen}, {ter Veen}, {Verbiest}, {Vermeulen}, {Vermaas}, {Vocks}, {Vogt}, {de Vos}, {van der Wal}, {van Weeren}, {Weggemans}, {Weltevrede}, {White}, {Wijnholds}, {Wilhelmsson}, {Wucknitz}, {Yatawatta}, {Zarka}, {Zensus}, \& {van Zwieten}}]{2013A&A...556A...2V}
{van Haarlem}, M.~P., {Wise}, M.~W., {Gunst}, A.~W., {et~al.} 2013, \aap, 556, A2

\bibitem[{{Walker} {et~al.}(2020){Walker}, {Hall}, {Allen}, {Green}, {Seitzer}, {Tyson}, {Bauer}, {Krafton}, {Lowenthal}, {Parriott}, {Puxley}, {Abbott}, {Bakos}, {Barentine}, {Bassa}, {Blakeslee}, {Bradshaw}, {Cooke}, {Devost}, {Galad{\'\i}-Enr{\'\i}quez}, {Haase}, {Hainaut}, {Heathcote}, {Jah}, {Krantz}, {Kucharski}, {McDowell}, {Mr{\'o}z}, {Otarola}, {Pearce}, {Rawls}, {Saunders}, {Seaman}, {Siminski}, {Snyder}, {Storrie-Lombardi}, {Tregloan-Reed}, {Wainscoat}, {Williams}, \& {Yoachim}}]{2020BAAS...52.0206W}
{Walker}, C., {Hall}, J., {Allen}, L., {et~al.} 2020, in Bulletin of the American Astronomical Society, Vol.~52, 0206

\bibitem[{{Wayth} {et~al.}(2022){Wayth}, {Sokolowski}, {Broderick}, {Tingay}, {Bhushan}, {Booler}, {Chiello}, {Davidson}, {Emrich}, {Juswardy}, {Kenney}, {Macario}, {Magro}, {Mattana}, {Minchin}, {Monari}, {McPhail}, {Perini}, {Pupillo}, {Schiaffino}, {Subrahmanyan}, {van Es}, {Walker}, \& {Waterson}}]{2022JATIS...8a1010W}
{Wayth}, R., {Sokolowski}, M., {Broderick}, J., {et~al.} 2022, Journal of Astronomical Telescopes, Instruments, and Systems, 8, 011010

\bibitem[{{White} {et~al.}(2023){White}, {Yu}, {Chen}, {Gary}, {Mondal}, {Fleishman}, {Nita}, {Bastian}, {Saint-Hilaire}, {McTiernan}, \& {Chen}}]{2023BAAS...55c.429W}
{White}, S., {Yu}, S., {Chen}, B., {et~al.} 2023, in Bulletin of the American Astronomical Society, Vol.~55, 429

\bibitem[{{White}(2024)}]{2024arXiv240500959W}
{White}, S.~M. 2024, arXiv e-prints, arXiv:2405.00959

\bibitem[{Wildemeersch \& Fortuny-Guasch(2010)}]{wildemeersch2010radio}
Wildemeersch, M. \& Fortuny-Guasch, J. 2010, EC Joint Research Centre, Security Tech. Assessment Unit, Tech. Rep, 50

\bibitem[{{Wilensky} {et~al.}(2020){Wilensky}, {Barry}, {Morales}, {Hazelton}, \& {Byrne}}]{2020MNRAS.498..265W}
{Wilensky}, M.~J., {Barry}, N., {Morales}, M.~F., {Hazelton}, B.~J., \& {Byrne}, R. 2020, \mnras, 498, 265

\bibitem[{Zarka(2024)}]{zarka2024star}
Zarka, P. 2024, arXiv preprint arXiv:2409.16038

\bibitem[{Zarka {et~al.}(2020)Zarka, Denis, Tagger, Girard, Coffre, Dumez-Viou, Taffoureau, Charrier, Bondonneau, Briand, {et~al.}}]{zarka2020low}
Zarka, P., Denis, L., Tagger, M., {et~al.} 2020, in URSI GASS 2020

\bibitem[{{Zarka} {et~al.}(2012){Zarka}, {Girard}, {Tagger}, \& {Denis}}]{2012sf2a.conf..687Z}
{Zarka}, P., {Girard}, J.~N., {Tagger}, M., \& {Denis}, L. 2012, in SF2A-2012: Proceedings of the Annual meeting of the French Society of Astronomy and Astrophysics, ed. S.~{Boissier}, P.~{de Laverny}, N.~{Nardetto}, R.~{Samadi}, D.~{Valls-Gabaud}, \& H.~{Wozniak}, 687--694

\bibitem[{{Zarka} {et~al.}(2015){Zarka}, {Lazio}, \& {Hallinan}}]{2015aska.confE.120Z}
{Zarka}, P., {Lazio}, J., \& {Hallinan}, G. 2015, in Advancing Astrophysics with the Square Kilometre Array (AASKA14), 120

\bibitem[{{Zucca} {et~al.}(2012){Zucca}, {Carley}, {McCauley}, {Gallagher}, {Monstein}, \& {McAteer}}]{2012SoPh..280..591Z}
{Zucca}, P., {Carley}, E.~P., {McCauley}, J., {et~al.} 2012, \solphys, 280, 591

\end{thebibliography}

\begin{appendix} 
\section{Summary of Starlink satellite detections with NenuFAR}
\longtab[1]{
\begin{longtable}{l c l c c c}
\caption{Summary of Starlink satellite detections with NenuFAR. The "Stokes I flux" and "Stokes V/I" were integrated over the 54-66 MHz band and 1 second. The “Stokes I flux” represents the actual flux density, corrected for primary beam effects. The “Peak Frequency” column indicates the frequency at which the maximum flux density was observed for each detection.}
\label{tab:passes}\\
\hline
\hline
Satellite Name & NORAD ID & Time of Detection & Stokes I Flux (Jy) & Stokes V/I & Peak Frequency (MHz) \\
\hline
\endfirsthead
\caption{Continued.} \\
\hline
Satellite Name & NORAD ID & Time of Detection & Stokes I Flux (Jy) & Stokes V/I & Peak Frequency (MHz) \\
\hline
\endhead
\hline
\endfoot
\hline
\endlastfoot
STARLINK-1073  &  44914  &  2024-06-20 04:33:23  & 10.24 & 0.64 & 56.34 \\
STARLINK-1133  &  45064  &  2024-06-22 03:12:07  & 10.76 & 0.58 & 60.10 \\
STARLINK-1593  &  46119  &  2024-06-22 01:49:48  & 20.75 & 0.42 & 54.37 \\
STARLINK-2127  &  47403  &  2024-06-24 05:47:50  & 9.03 & 0.63 & 55.06 \\
STARLINK-2067  &  47674  &  2024-06-25 07:02:09  & 9.45 & 0.71 & 54.17 \\
STARLINK-2078  &  47675  &  2024-06-25 06:52:33  & 6.91 & 0.73 & 58.32 \\
STARLINK-2369  &  47909  &  2024-06-21 03:52:59  & 6.79 & 0.64 & 54.07 \\
STARLINK-2254  &  48005  &  2024-06-21 04:34:52  & 6.95 & 0.67 & 55.26 \\
STARLINK-2623  &  48368  &  2024-06-21 04:22:52  & 6.95 & 0.68 & 60.01 \\
STARLINK-2621  &  48370  &  2024-06-20 04:51:11  & 5.38 & 0.62 & 54.76 \\
STARLINK-3180  &  51137  &  2024-06-21 05:43:35  & 5.77 & 0.71 & 63.77 \\
STARLINK-3514  &  51731  &  2024-06-25 05:47:17  & 9.42 & 0.72 & 54.17 \\
STARLINK-3502  &  51746  &  2024-06-24 05:16:24  & 36.34 & 0.67 & 54.27 \\
STARLINK-3597  &  51898  &  2024-07-03 00:43:23  & 3.00 & 0.73 & 61.87 \\
STARLINK-3648  &  51993  &  2024-06-25 05:54:45  & 9.22 & 0.75 & 58.32 \\
STARLINK-4012  &  52602  &  2024-06-21 04:27:07  & 7.99 & 0.65 & 54.47 \\
STARLINK-4046  &  52842  &  2024-06-25 05:31:32  & 4.90 & 0.71 & 58.91 \\
STARLINK-4301  &  52997  &  2024-06-25 06:18:11  & 7.09 & 0.75 & 54.96 \\
STARLINK-4328  &  53071  &  2024-07-03 04:58:31  & 2.68 & 0.76 & 54.37 \\
STARLINK-4799  &  53860  &  2024-07-03 00:36:50  & 8.56 & 0.73 & 54.37 \\
STARLINK-4788  &  53863  &  2024-07-03 00:25:50  & 3.21 & 0.74 & 64.84 \\
STARLINK-4775  &  53934  &  2024-06-25 04:19:50  & 10.09 & 0.67 & 54.96 \\
STARLINK-5226  &  54056  &  2024-06-21 03:52:37  & 4.81 & 0.63 & 62.19 \\
STARLINK-5250  &  54200  &  2024-07-03 05:14:01  & 2.15 & 0.75 & 56.94 \\
STARLINK-5331  &  55280  &  2024-06-20 04:50:28  & 5.53 & 0.64 & 55.16 \\
STARLINK-5286  &  55294  &  2024-06-20 04:50:14  & 5.38 & 0.67 & 65.63 \\
STARLINK-30296  &  57671  &  2024-06-25 04:45:19  & 22.96 & 0.69 & 63.85 \\
STARLINK-30313  &  57676  &  2024-06-25 04:59:53  & 44.98 & 0.66 & 60.29 \\
STARLINK-30297  &  57682  &  2024-06-25 04:52:55  & 16.23 & 0.67 & 60.29 \\
STARLINK-30295  &  57683  &  2024-06-24 04:58:37  & 22.46 & 0.57 & 57.83 \\
STARLINK-30416  &  57847  &  2024-06-25 04:46:05  & 42.06 & 0.70 & 58.52 \\
STARLINK-30416  &  57847  &  2024-06-24 05:13:28  & 25.68 & 0.64 & 58.03 \\
STARLINK-30430  &  57849  &  2024-06-24 05:13:59  & 28.67 & 0.63 & 64.56 \\
STARLINK-30430  &  57849  &  2024-06-25 04:46:35  & 43.54 & 0.55 & 62.27 \\
STARLINK-30711  &  57940  &  2024-06-20 17:55:27  & 73.55 & 0.46 & 55.36 \\
STARLINK-30455  &  57942  &  2024-06-21 17:35:06  & 45.49 & 0.36 & 55.36 \\
STARLINK-30455  &  57942  &  2024-06-20 18:02:24  & 98.85 & 0.48 & 55.36 \\
STARLINK-30456  &  57943  &  2024-06-22 17:14:41  & 47.79 & 0.49 & 64.44 \\
STARLINK-30704  &  57946  &  2024-06-21 18:25:05  & 47.45 & 0.40 & 55.36 \\
STARLINK-30713  &  57952  &  2024-06-21 17:35:35  & 29.33 & 0.45 & 55.36 \\
STARLINK-30713  &  57952  &  2024-06-20 18:02:56  & 59.92 & 0.49 & 55.36 \\
STARLINK-30562  &  58033  &  2024-06-20 18:09:43  & 27.03 & 0.52 & 55.36 \\
STARLINK-30533  &  58037  &  2024-06-21 17:53:41  & 40.42 & 0.45 & 55.36 \\
STARLINK-30559  &  58039  &  2024-06-20 18:19:18  & 26.96 & 0.55 & 55.36 \\
STARLINK-30555  &  58040  &  2024-06-20 18:14:31  & 61.13 & 0.45 & 55.36 \\
STARLINK-30529  &  58047  &  2024-06-21 17:44:04  & 30.52 & 0.44 & 55.36 \\
STARLINK-30595  &  58111  &  2024-06-21 03:17:26  & 44.77 & 0.59 & 61.40 \\
STARLINK-30595  &  58111  &  2024-06-22 02:49:53  & 22.73 & 0.52 & 59.90 \\
STARLINK-30353  &  58113  &  2024-06-20 03:37:17  & 37.33 & 0.61 & 61.08 \\
STARLINK-30572  &  58114  &  2024-06-22 02:56:52  & 34.41 & 0.48 & 62.27 \\
STARLINK-30784  &  58156  &  2024-06-22 01:43:09  & 21.81 & 0.51 & 64.64 \\
STARLINK-30814  &  58165  &  2024-06-25 05:29:38  & 44.99 & 0.53 & 60.69 \\
STARLINK-30814  &  58165  &  2024-06-24 05:56:57  & 26.94 & 0.60 & 57.83 \\
STARLINK-30811  &  58167  &  2024-06-25 05:37:08  & 35.96 & 0.61 & 58.52 \\
STARLINK-30812  &  58168  &  2024-06-25 05:36:37  & 14.61 & 0.72 & 62.07 \\
STARLINK-30771  &  58169  &  2024-06-25 05:30:09  & 14.72 & 0.70 & 62.66 \\
STARLINK-30771  &  58169  &  2024-06-24 05:57:27  & 28.07 & 0.63 & 62.38 \\
STARLINK-30764  &  58170  &  2024-06-24 05:49:59  & 35.35 & 0.50 & 62.38 \\
STARLINK-30938  &  58389  &  2024-06-20 16:48:04  & 104.22 & 0.40 & 55.36 \\
STARLINK-31012  &  58531  &  2024-06-20 16:02:58  & 149.86 & 0.45 & 57.53 \\
STARLINK-31063  &  58547  &  2024-06-20 20:29:35  & 94.78 & 0.37 & 60.89 \\
STARLINK-31031  &  58711  &  2024-06-22 02:55:03  & 90.72 & 0.50 & 62.47 \\
STARLINK-31098  &  58716  &  2024-06-20 03:49:50  & 23.76 & 0.64 & 60.69 \\
STARLINK-31093  &  58717  &  2024-06-22 02:59:52  & 68.36 & 0.46 & 63.85 \\
STARLINK-31097  &  58718  &  2024-06-20 03:45:02  & 55.56 & 0.46 & 60.89 \\
STARLINK-31051  &  58719  &  2024-06-20 03:40:13  & 69.17 & 0.57 & 61.28 \\
STARLINK-31027  &  58720  &  2024-06-20 03:35:25  & 22.53 & 0.67 & 64.84 \\
STARLINK-30972  &  58721  &  2024-06-21 03:25:47  & 38.22 & 0.65 & 62.98 \\
STARLINK-31040  &  58722  &  2024-06-21 03:19:23  & 63.09 & 0.53 & 63.57 \\
STARLINK-30893  &  58724  &  2024-06-21 03:14:34  & 46.57 & 0.55 & 63.97 \\
STARLINK-31099  &  58781  &  2024-06-25 06:22:35  & 12.94 & 0.68 & 58.12 \\
STARLINK-31047  &  58783  &  2024-06-25 06:08:09  & 21.75 & 0.65 & 57.92 \\
STARLINK-31039  &  58784  &  2024-06-25 06:17:46  & 107.87 & 0.58 & 61.08 \\
STARLINK-31034  &  58785  &  2024-06-25 04:31:59  & 19.68 & 0.54 & 57.92 \\
STARLINK-31034  &  58785  &  2024-06-06 16:29:50  & 203.33 & 0.48 & 58.32 \\
STARLINK-31024  &  58786  &  2024-06-25 06:12:58  & 23.14 & 0.62 & 59.50 \\
STARLINK-31233  &  58826  &  2024-06-25 05:14:29  & 93.16 & 0.41 & 60.49 \\
STARLINK-31246  &  58835  &  2024-06-25 05:15:00  & 42.76 & 0.49 & 60.10 \\
STARLINK-31251  &  58836  &  2024-06-24 05:35:19  & 43.76 & 0.64 & 58.03 \\
STARLINK-31251  &  58836  &  2024-06-25 05:08:00  & 121.56 & 0.29 & 58.12 \\
STARLINK-31249  &  58838  &  2024-06-25 05:07:31  & 59.46 & 0.45 & 60.29 \\
STARLINK-31249  &  58838  &  2024-06-24 05:34:48  & 63.84 & 0.57 & 58.23 \\
STARLINK-31250  &  58839  &  2024-06-24 05:28:20  & 54.61 & 0.50 & 60.60 \\
STARLINK-31248  &  58840  &  2024-06-22 01:20:58  & 106.89 & 0.62 & 64.64 \\
STARLINK-31236  &  58841  &  2024-06-24 05:27:50  & 60.85 & 0.41 & 58.03 \\
STARLINK-31228  &  58842  &  2024-06-22 01:20:27  & 111.59 & 0.53 & 64.44 \\
STARLINK-31262  &  58847  &  2024-06-22 02:39:41  & 43.13 & 0.51 & 61.48 \\
STARLINK-31323  &  58882  &  2024-06-25 07:42:10  & 50.78 & 0.51 & 58.12 \\
STARLINK-31209  &  58883  &  2024-06-25 07:41:41  & 65.20 & 0.53 & 57.92 \\
STARLINK-31320  &  58886  &  2024-06-25 07:35:12  & 104.26 & 0.47 & 57.73 \\
STARLINK-31230  &  58890  &  2024-07-04 04:18:26  & 93.04 & 0.42 & 61.48 \\
STARLINK-31390  &  58937  &  2024-06-25 06:36:16  & 124.58 & 0.41 & 58.12 \\
STARLINK-31377  &  58938  &  2024-06-25 06:35:46  & 74.80 & 0.56 & 58.32 \\
STARLINK-31371  &  58940  &  2024-06-25 06:28:47  & 11.80 & 0.60 & 58.52 \\
STARLINK-31385  &  58985  &  2024-06-25 05:30:20  & 68.11 & 0.48 & 60.69 \\
STARLINK-31591  &  59207  &  2024-06-20 03:29:54  & 60.23 & 0.47 & 61.68 \\
STARLINK-31520  &  59210  &  2024-06-20 03:29:24  & 43.03 & 0.47 & 62.86 \\
STARLINK-31523  &  59220  &  2024-06-20 03:36:20  & 27.51 & 0.59 & 60.49 \\
STARLINK-31516  &  59221  &  2024-06-22 02:47:35  & 39.78 & 0.44 & 61.68 \\
STARLINK-31629  &  59255  &  2024-07-03 00:12:42  & 55.35 & 0.45 & 61.48 \\
STARLINK-31629  &  59255  &  2024-06-21 04:39:50  & 21.34 & 0.45 & 58.03 \\
STARLINK-31598  &  59263  &  2024-06-21 04:32:53  & 54.00 & 0.51 & 60.80 \\
STARLINK-31598  &  59263  &  2024-07-03 00:05:44  & 10.94 & 0.50 & 61.28 \\
STARLINK-31645  &  59264  &  2024-06-20 04:57:24  & 20.32 & 0.53 & 60.69 \\
STARLINK-31623  &  59272  &  2024-06-24 03:26:38  & 34.86 & 0.59 & 57.83 \\
STARLINK-31641  &  59374  &  2024-06-25 03:23:57  & 71.00 & 0.49 & 62.86 \\
STARLINK-31430  &  59375  &  2024-06-24 03:55:53  & 52.16 & 0.48 & 60.80 \\
STARLINK-31695  &  59376  &  2024-06-21 05:23:08  & 36.01 & 0.63 & 57.83 \\
STARLINK-31716  &  59377  &  2024-06-25 03:24:24  & 86.19 & 0.52 & 63.06 \\
STARLINK-31710  &  59378  &  2024-07-03 00:42:04  & 75.81 & 0.61 & 62.47 \\
STARLINK-31698  &  59379  &  2024-06-21 05:22:34  & 36.09 & 0.63 & 60.60 \\
STARLINK-31680  &  59380  &  2024-06-24 03:48:28  & 113.41 & 0.42 & 62.38 \\
STARLINK-31713  &  59381  &  2024-06-25 03:30:51  & 92.47 & 0.58 & 62.66 \\
STARLINK-31705  &  59382  &  2024-06-21 05:15:38  & 51.87 & 0.52 & 59.81 \\
STARLINK-31643  &  59388  &  2024-07-03 00:41:37  & 21.26 & 0.55 & 65.03 \\
STARLINK-31647  &  59389  &  2024-06-24 03:48:58  & 24.59 & 0.65 & 57.83 \\
STARLINK-31649  &  59390  &  2024-06-21 05:16:10  & 80.54 & 0.44 & 60.80 \\
STARLINK-31443  &  59393  &  2024-07-03 04:58:42  & 82.17 & 0.38 & 61.68 \\
STARLINK-31443  &  59393  &  2024-07-04 04:20:14  & 63.22 & 0.44 & 63.45 \\
STARLINK-31443  &  59393  &  2024-07-03 00:05:33  & 39.40 & 0.54 & 64.64 \\
STARLINK-31443  &  59393  &  2024-06-21 04:32:41  & 22.64 & 0.66 & 61.79 \\
STARLINK-11085  &  59425  &  2024-06-25 04:26:34  & 17.54 & 0.64 & 63.85 \\
STARLINK-31740  &  59433  &  2024-06-25 05:28:16  & 11.11 & 0.58 & 58.52 \\
STARLINK-31580  &  59434  &  2024-06-24 05:59:48  & 75.13 & 0.43 & 60.60 \\
STARLINK-31843  &  59722  &  2024-06-22 02:03:32  & 65.19 & 0.52 & 62.07 \\
STARLINK-11118  &  59754  &  2024-06-20 18:05:49  & 27.10 & 0.51 & 55.36 \\
STARLINK-11118  &  59754  &  2024-06-20 13:20:06  & 133.55 & 0.28 & 57.92 \\
STARLINK-11132  &  59755  &  2024-06-20 18:02:49  & 47.11 & 0.50 & 55.36 \\
STARLINK-31826  &  59768  &  2024-06-21 17:52:12  & 62.22 & 0.47 & 55.36 \\
STARLINK-31826  &  59768  &  2024-06-20 13:37:45  & 261.41 & 0.28 & 61.08 \\
STARLINK-31829  &  59770  &  2024-06-22 17:34:37  & 38.31 & 0.47 & 62.27 \\
STARLINK-11124  &  59952  &  2024-06-20 16:38:39  & 50.54 & 0.40 & 55.36 \\
STARLINK-11120  &  60042  &  2024-06-20 17:30:48  & 183.86 & 0.36 & 55.36 \\
STARLINK-11086  &  60043  &  2024-06-20 17:27:39  & 14.77 & 0.26 & 55.36 \\                  
\end{longtable}
}

\end{appendix}

\end{document}